\documentclass[printer]{aa}

\bibpunct{(}{)}{;}{a}{}{,}

\usepackage{graphicx}
\usepackage{txfonts}
\usepackage{soul}
\usepackage{txfonts}
\usepackage{color}
\usepackage{ulem}
\usepackage{lscape}
\usepackage{longtable}
\usepackage{rotating}
\usepackage{pdflscape}

\newcommand{\hii}{H\textsc{ii}}
\def\ks{km s$^{-1}$}

\def\cm3{cm$^{-3}$}

\def\2{$^{12}$CO}
\def\3{$^{13}$CO}
\def\8{C$^{18}$O}

\def\cm2{cm$^{-2}$}

\begin{document}

\title{Sulfur-bearing molecules in a sample of early star-forming cores}

\author{ N. C. Martinez\inst{1,2}
\and S. Paron \inst{1}
\and M. E. Ortega\inst{1}
\and A. Petriella\inst{1}
\and A. Álamo\inst{3,1}
\and M. Brook \inst{2,1}
\and C. Carballo\inst{4,1}
\and T. Heberling\inst{1}
}
\institute{CONICET - Universidad de Buenos Aires. Instituto de Astronom\'{\i}a y F\'{\i}sica del Espacio, Buenos Aires, Argentina\\
             \email{nmartinez@iafe.uba.ar}
\and Universidad de Buenos Aires, Facultad de Ciencias Exactas y Naturales, Buenos Aires, Argentina
\and Instituto Superior de Formación Docente N$^{\rm o}$41, Adrogué, Argentina
\and Universidad Nacional de La Plata, Facultad de Ciencias Astronómicas y Geofísicas, La Plata, Argentina
}

\offprints{N. C. Martinez}

   \date{Received <date>; Accepted <date>}

\abstract{}{The sulfur content in dense molecular regions is highly depleted in comparison to diffuse clouds. Given
that the reason of this phenomenon is unclear, it is necessary to carry on observational studies of sulfur-bearing species towards 
dense regions, mainly at early evolutive stages. In this line, analysis of sulfur bearing molecules in a large sample of dense starless molecular cores is of great importance to help to uncover the early sulfur chemistry in such regions. 
}
{From the Atacama Large Millimeter Array (ALMA) data archive, we selected a project in Band 7 (275--373 GHz), which contains the emission of several sulfur-bearing species, done towards a sample of 37 dense cores embedded in the most massive infrared-quiet molecular clumps from the ATLASGAL survey. Lines of $^{34}$SO, SO$_{2}$, NS, SO, SO$^{+}$, and H$_{2}$CS were analyzed and column densities of each molecular species were obtained. From the continuum emission, and two CH$_{3}$OH lines, the 37 cores were characterized in density and temperature, and the corresponding H$_{2}$ column densities were derived. The abundances of such sulfur-bearing species were derived and studied.}{
We find that the abundances of the
analyzed sulfur-bearing species increase with the growth of the gas temperature. From the correlation between abundances and temperature, we suggest that the
chemistry involved in the formation of each of the analyzed molecule may have a similar dependence with T$_{k}$ in the range 20 to 100 K. Additionally we find that the comparisons among abundances are, in general, highly correlated. Taking into account that such correlation  decreases in more evolved sources, we suggest that the sulfur-bearing species here analyzed should have a similar
chemical origin. Our observational results give support to the use of the X(SO$_{2}$)/X(SO) ratio as a chemical clock of molecular cores. From the line widths of the molecular lines we point out that molecules with oxygen content ($^{34}$SO, SO$_{2}$, SO, and SO$^{+}$) may be associated with warmer and more turbulent gas than the other ones. H$_{2}$CS and NS are associated with more quiescent gas, probably in the external envelopes of the cores, tracing similar physical and chemical conditions. This work not only complement, with a large sample of sources, recent similar works done towards more evolved sources, but also it gives quantitative information about abundances that could be useful
in chemical models pointing to explain the sulfur chemistry in the interstellar medium.
}{}

\titlerunning{Sulfur-bearing molecules in star-forming regions}
\authorrunning{N. C. Martinez et al.}


\maketitle

\section{Introduction}

Sulfur (S) is the tenth most abundant element in the Universe and the sixth most important biogenic one \citep{fontani23}; indeed, it is the fifth most abundant element on our planet. Even though the chemistry related to the sulfur in the interstellar medium (ISM) has been continuously investigated since many years ago (e.g. \citealt{pineau86,turner95,wak04,goico06, vastel18,fontani23}), it still is far from being well understood. 

One of the unsolved main problems regarding the sulfur chemistry in the ISM is that its content in dense molecular regions is highly depleted compared to diffuse clouds \citep{laas19,riviere19}. The abundance of sulfur in the gas-phase measured through molecular emission in dense areas of the ISM is estimated to be at most a few percent of the cosmic value \citep{woods15,fontani23}. \citet{hily22} found that, towards early-type cores, the sulfur abundance is close to the cosmic abundance. The authors proposed that sulfur is not depleted and is atomic in such environments, which is in agreement with some chemical models. Then, according to the authors, more chemically evolved cores show sulfur depletion by factors up to 100 in their densest parts.  

The principal hypothesis aimed to explain the mentioned depletion is that most of the sulfur should be locked into dust grains in the densest regions of the ISM \citep{laas19}. This generally involves species in ices, such as organo-sulfur molecules, sulfur chains, or rings such as S$_{8}$ \citep{shing20}. However, only OCS has been detected on ices, and only upper limits have been obtained for H$_{2}$S, leaving unknown the main carrier of sulfur (\citealt{hily22}, and references therein).

Recently, \citet{fuente23} proposed that the sulfur depletion in the gas-phase might depend on the particularities of each environment. The
mechanisms responsible for this dependence are not known in full detail. The authors proposed that the influence of the grain charges on the chemistry could be one of the causes. Additionally, shocks associated with massive star formation could erode the grain cores, which can contribute to enhancing the sulfur elemental abundance in the hosting molecular cloud. 

Despite the considerable efforts performed to determine the location of the missing interstellar sulfur, the primary carrier of such an element is still currently unknown in both gas and solid-phase \citep{artur23}. Hence, more studies of sulfur-bearing species towards dense regions of the ISM are required. 

Additionally, as \citet{fontani23} point out, another unresolved problem regarding the S chemistry is that the formation of some S-bearing molecules detected in the gas-phase is not evident. For instance, some basic reactions that initiate sulfur chemistry are highly endothermic, requiring temperatures much higher than those typically found in cold, dense molecular cores and clumps. Therefore, further studies on the detection of sulfur-bearing species in molecular cores are necessary.

In conclusion, abundance measurements of S-bearing molecules and comparisons among them towards different interstellar environments are essential for understanding sulfur chemistry. In particular, cores at the early stages of star-forming evolution should be of notable interest to constrain the early sulfur chemistry in such environments. Hence, using data from the Atacama Large Millimeter Array (ALMA) data archive (see Sect.\,\ref{sectdata}), in this paper we present a study of several sulfur species towards 37 dense cores embedded in the most massive infrared-quiet molecular clumps from the ATLASGAL survey.

\section{Source sample and data}
\label{sectdata}

To carry on this work we first looked in the ALMA database for data containing several lines of sulfur-bearing species. Data acquired in the ALMA Band 7 fulfill this requirement.
We found that project 2017.1.00914 (PI: Csengeri) consists of observations in Band 7 towards massive infrared quiet clumps in the inner Galaxy, i.e. ATLASGAL sources likely containing 
fragments, such as molecular cores probably at early stages of star formation. Following a thorough examination of the data cubes and a survey of sulfur-bearing molecular species across the entire dataset, we selected the molecular lines listed in Table\,\ref{transitions}.

From the continuum maps at 0.87 mm obtained towards each ATLASGAL source, molecular cores embedded in such clumps were identified. In cases where multiple cores are present in the clump, the most intense one was selected. The ATLASGAL source names and the coordinates of the selected and analyzed molecular cores are presented in Table\,\ref{sampleoriginal}. The systemic velocities (v$_{\rm LSR}$), needed to identify the molecular lines, were obtained from the cataloged velocity of the ATLASGAL source, and slightly corrected when necessary, using the most intense and ubiquitous line among the selected molecules, thioformaldehyde (H$_{2}$CS). Distances were obtained from the catalogs in Vizier `ATLASGAL cold high-mass clumps with NH$_{3}$' \citep{wie12} and `Complete sample of Galactic clump properties' \citep{urqu18}.

\begin{table}[h]
\centering
\small
\caption{Analyzed sulfur-bearing molecular lines.}
\label{transitions}
\begin{tabular}{llc}
\hline
\hline
Molecule & Transition & Rest Freq. (GHz)   \\
\hline
 & &   \\[-1.8ex]
$^{34}$SO                   & 7(8)--6(7)            & 333.900      \\
SO$_{2}$ v=0                & 8(2,6)--7(1,7)        & 334.673     \\
NS v=0                      & 15/2--13/2            & 346.221     \\
SO v=0 $^{3}$$\Sigma$       & 9(8)--8(7)            & 346.528     \\
SO$^{+}$                    & 15/2--13/2 (1/2) l=f  & 348.115    \\
H$_{2}$CS                   &  10(1,9)--9(1,8)      & 348.534     \\
\hline
 & &   \\[-1.8ex]
\end{tabular}
\end{table}

\begin{table*}[h]
\centering
\caption{ATLASGAL sources sample and selected cores.}
\label{sampleoriginal}
\small
\begin{tabular}{lccccc}
\hline
\hline
\#   & ATLASGAL  & \multicolumn{2}{c}{Core coordinates} & v$_{\rm LSR}$ & Dist.    \\  
    &    source      & $\alpha$ (J2000) & $\delta$ (J2000)  &  (km s$^{-1}$) & (kpc)  \\
\hline
1 & 006.1879$-$0.3586	&	18:01:02.075 & -23.47.11.525	&	-33.8 &  N/A$^{(*)}$ \\
2 & 008.6702$-$0.3557	&	18:06:19.067 & -21.37.31.483	&	35.0  & 4.4 \\
3 & 008.6834$-$0.3675	&	18:06:23.496 & -21.37.10.784	&	39.0  & 4.4 \\
4 & 011.9176$-$0.6133	&	18:13:58.110 & -18.54.20.328	&	35.7  & 3.7 \\
5 & 012.4166+0.5053	    &	18:10:51.328 & -17.55.46.940	&	17.0  & 2.3 \\
6 & 012.6791$-$0.1810	&	18:13:54.722 & -18.01.46.484	&	55.5  & 4.6 \\
7 & 012.8882+0.4897	    &	18:11:51.461 & -17.31.28.831	&	31.0  & 2.4 \\
8 & 014.2273$-$0.5109	&	18:18:12.308 & -16.49.27.843	&	19.7  & 2.3 \\
9 & 014.3314$-$0.6446	&	18:18:54.562 & -16.47.50.178	&	22.5  & 1.1 \\
10 & 014.6323$-$0.5763	&	18:19:15.199 & -16.30.05.060	&	17.5  & 2.2 \\
11 & 016.5850$-$0.0503	&	18:21:09.070 & -14.31.47.976	&	59.7  & 4.7 \\
12 & 019.0762$-$0.2876	&	18:26:48.831 & -12.26.24.924	&	65.5  & 4.5 \\
13 & 019.8832$-$0.5347	&	18:29:14.407 & -11.50.23.130	&	44.0  & 3.3 \\
14 & 023.2056$-$0.3772	&	18:34:55.198 & -08.49.15.143	&	77.5  & 4.6 \\ 
15 & 028.3967+0.0805 	&	18:42:52.053 & -03.59.54.356	&	78.5  & 4.5 \\
16 & 028.8311$-$0.2527	&	18:44:51.065 & -03.45.49.044	&	86.8  & 4.9 \\
17 & 037.4301+1.5192	&	18:54:14.257 &  04.41.40.259	&	44.0  & 2.4 \\
18 & 045.4656+0.0457 	&	19:14:25.664 &  11.09.25.100	&	64.0  & 7.8 \\
19 & 305.3666+0.2130	&	13:12:36.168 & -62.33.34.492	&	-34.6 & 4.0 \\
20 & 305.7994$-$0.2445	&	13:16:43.215 & -62.58.32.589	&	-32.5 & 4.0 \\
21 & 305.8874+0.0164	&	13:17:15.419 & -62.42.23.306	&	-33.0 & 4.0 \\
22 & 310.1440+0.7592	&	13:51:58.278 & -61.15.41.527	&	-56.0 & 5.4 \\
23 & 322.9314+1.3933	&	15:20:21.031 & -55.35.13.555	&	-39.0 & 2.7 \\
24 & 327.3039$-$0.5775	&	15:53:10.919 & -54.36.46.828	&	-47.0 & 3.0 \\
25 & 327.5669$-$0.8522	&	15:55:49.503 & -54.39.16.426	&	-35.0 & 2.3 \\
26 & 329.0656$-$0.3076	&	16:01:09.866 & -53.16.02.484	&	-42.6 & 2.9 \\
27 & 332.6938$-$0.6125	&	16:19:51.586 & -51.01.33.266	&	-48.6 & 3.1 \\
28 & 335.0604$-$0.4269	&	16:29:23.178 & -49.12.29.333	&	-40.0 & 2.8 \\
29 & 337.4052$-$0.4024	&	16:38:50.527 & -47.28.01.168	&	-41.3 & 3.1 \\
30 & 337.9154$-$0.4773	&	16:41:10.489 & -47.08.02.881	&	-39.0 & 3.0 \\
31 & 338.9188+0.5494	&	16:40:34.005 & -45.42.07.903	&	-63.0 & 4.2 \\
32 & 345.0029$-$0.2241	&	17:05:10.903 & -41.29.06.724	&	-27.1 & 2.8 \\
33 & 348.7007$-$1.0427	&	17:20:03.964 & -38.58.30.521	&	-13.5 & 2.0 \\
34 & 351.4659+0.6821	&	17:20:52.937 & -35.43.16.485	&	-3.4  & 1.3\\
35 & 351.7437$-$0.5768	&	17:26:47.503 & -36.12.11.413	&	-2.8  & 1.3 \\
36 & 351.7853$-$0.5134	&	17:26:38.892 & -36.08.08.759	&	-2.6  & 1.3 \\
37 & 359.6158$-$0.2429	&	17:45:39.055 & -29.23.30.517	&	19.0  & 7.7 \\
\hline
\multicolumn{6}{l}{\small (*)  No kinematic distance solution obtained from galactic rotation models. } \\ 
\end{tabular}
\end{table*}

The data cubes from the project 2017.1.00914 (PI: Csengeri, T.; Band 7) were obtained from the ALMA Science Archive\footnote{http://almascience.eso.org/aq/}. The single-pointing observations for the 37 targets were carried out using the following telescope configuration: L5BL/L80BL(m): 8.7/27.5 in the 7 m array. The observed
frequency range and the spectral resolution are 333.3–349.1 GHz and 1.1 MHz, respectively. The angular resolution goes from 3$\farcs$3 to 3$\farcs$8 for the whole sample. The continuum and line (each 10 km\,s$^{-1}$) sensitivity are about 1.2 and 30~mJy beam$^{-1}$.

It is important to remark that even though the data passed the QA2 quality level, which assures a reliable calibration for a ``science ready'' data, the automatic pipeline imaging process may give rise to a clean image with some artifacts. For example, an inappropriate setting of the parameters of the {\it clean} task in CASA could generate artificial dips in the spectra. Thus, we reprocessed some raw data using CASA 4.5.1 and 4.7.2 versions and the calibration pipeline scripts. Particular care was taken with the different parameters of the task {\it clean}. The images and spectra from our data reprocessing, following multiple runs of the {\it clean} task with varied parameters, closely resembled the archival data. Consequently, we chose to rely on the archival data.

In some cases, when it was appreciated that the spectral lines were mounted on some continuum emission, the task {\it imcontsub} in CASA was used to subtract it. A first-order polynomial was enough for an appropriate continuum subtraction after carefully selecting frequency ranges without molecular line emission in each spectral window.

\section{Results}

Once the cores within each ATLASGAL source have been identified, the spectra from each of them were extracted at their continuum peak position using a beam-size circular region.  It is important to mention that in all cases, the peaks of the molecular emissions coincide with those of the continuum emission.  After the six molecular lines were identified, Gaussian fittings to each line were applied to obtain the intensity peaks, the FWHM $\Delta$v, and the integrated line intensity (i.e. the line flux, $W$). In Appendix\,\ref{append1} all these parameters are presented in Tables\,\ref{param1} and \ref{param2}, and in Appendix\,\ref{example} spectra and Gaussian fittings of a particular core are shown as an example of this procedure. 

None of the analyzed lines showed signs of absorption or flattening, and all were well-fitted by Gaussians, indicating that opacity effects, if any, likely had little impact on our analysis.

Additionally, it is important to note that, in general, the lines present well defined Gaussian shapes, with no signature of large spectral wings. SO and SO$_{2}$ are the lines that more frequently present some signatures of such wings, and, to a lesser extent, NS and H$_{2}$CS. In the cases that the lines have spectral wings, only the main central component was fitted (see figures in Appendix\,\ref{example}). Such small spectral wings could be due to some dynamical process in the gas or to the presence of other molecular lines, probably COMs (not identified, and out of the scope of this work).

To estimate the column densities of each molecular species, assuming local thermodynamic equilibrium (LTE) and following \citet{artur23}, firstly we derive the column densities of the
upper levels, N$_{\rm u}$ (in cm$^{-2}$), of each transition from:

\begin{equation}
{\rm N_u = 2375 \times 10^6~\left(\frac{W}{1~Jy~kms^{-1}}\right)~\left(\frac{1~s^{-1}}{A_{ul}}\right)~\left(\frac{arcsec^2}{\Theta_{area}}\right)}
\label{coldensEq}
\end{equation}

\noindent where  W is the measured line flux in Jy beam$^{-1}$ km s$^{-1}$ (see Tables \ref{param1} and \ref{param2}), $A_{ul}$ the line Einstein coefficient (in s$^{-1}$) (see Table \ref{param}), $\rm \Theta_{area}$ is the area of integration in arcsec$^2$, which was taken equal to the size of the beam in all cases.

Then, the total column density of each molecule was obtained from:

\begin{equation}
{\rm N_{tot} = \frac{N_u~Q(T_{ex})~exp(E_{u}/T_{ex})}{g_u}}
\label{coldensEq}
\end{equation}

\noindent where ${\rm Q(T_{ex})}$ is the rotational partition function, ${\rm T_{ex}}$ the excitation temperature (in K), $g_{u}$ the upper-level degeneracy, and E$_{\rm up}$ (in K) the upper energy of the line transition.  The used parameters are shown in Table\,\ref{param}. A ${\rm T_{ex}}$ of 20 K was assumed for the excitation of all lines uniformly in the whole sample. The obtained column densities and their respective errors are presented in Table\,\ref{coldens}.
 
It is important to note that for these calculations it was assumed that all transitions are optically thin, supported by the lack of high opacity evidence, such as absorption dips or flattened components. However, given that we cannot completely rule out this phenomenon, $\tau$ values were obtained for all the lines in the whole sample of cores (see Appendix\,\ref{tauapp}). The optically thin assumption is supported by such obtained values.

\begin{table}[h]
\centering
\small
\caption{Molecular line parameters.}
\label{param}
\begin{tabular}{lcccc}
\hline
\hline
Line &    E$\rm_{u}$ (K) & $g_{u}$ & log($A_{ul}$) & $Q(20~{\rm K}$)  \\
\hline
    &                &         &               &                \\[-1.8ex]
$^{34}$SO 7(8)--6(7)               & 79.8  & 15  & -3.32 & 39.5   \\
SO$_{2}$ v=0 8(2,6)--7(1,7)        & 43.1  & 17  & -3.89 & 92.8    \\
NS v=0  15/2--13/2                 & 69.8  & 16  & -3.14 & 100.5 \\
SO v=0 $^{3}$$\Sigma$  9(8)--8(7)  & 78.7  & 19  & -3.26 & 38.8 \\
SO$^{+}$ 15/2--13/2 (1/2) l=f      & 70.2  & 16  & -3.64 & 35.3    \\
H$_{2}$CS 10(1,9)--9(1,8)          & 105.2 & 21  & -3.20 & 91.2 \\
\hline
 & &  &   &    \\[-1.8ex]
\multicolumn{5}{l}{\tiny Note: Parameters extracted from the Splatalogue Database for}\\ 
\multicolumn{5}{l}{\tiny Astronomical Spectroscopy using the Cologne Database for Molecular } \\
\multicolumn{5}{l}{\tiny Spectroscopy (CDMS; \citealt{muller05}). }\\ 
\multicolumn{5}{l}{\tiny References for the different molecular lines:}\\
\multicolumn{5}{l}{\tiny $^{34}$SO: \citet{tiemann74}; SO$_{2}$: \citet{lovas78} and \citet{belov98}; }\\
\multicolumn{5}{l}{\tiny SO: \citet{clark76}; NS: \citet{lee95};}\\
\multicolumn{5}{l}{\tiny SO$^{+}$: \citet{amano91}; H$_{2}$CS: \citet{maeda08}. }\\
\end{tabular}
\end{table}

\begin{table*}[h]
\centering
\caption{LTE column densities ($\times 10^{13}$ cm$^{-2}$).}
\label{coldens}
\begin{tabular}{lcccccccccccc}
\hline
\hline
Core&N(SO$^{+}$)&Error&N(H$_{2}$CS)&Error&N(NS)&Error&N(SO)&Error&N($^{34}$SO)&Error&N(SO$_{2}$)&Error \\
\hline
1&-&-&92.00&4.68&8.88&0.73&7.84&0.23&1.93&0.19&7.58&0.17\\
2&2.60&0.30&214.00&16.80&34.80&2.13&23.60&1.09&6.51&0.43&18.10&0.37\\
3&1.60&0.15&112.00&12.80&12.80&1.53&12.70&0.54&3.24&0.19&15.00&0.51\\
4&1.84&0.22&152.00&4.68&35.90&1.66&25.80&0.43&5.54&0.16&30.20&0.91\\
5&0.57&0.22&26.40&4.06&6.20&0.06&4.75&0.24&0.64&0.16&1.43&0.21\\
6&5.35&0.52&243.00&22.80&48.40&3.79&27.20&0.97&11.00&0.53&56.40&1.70\\
7&15.40&0.82&527.00&37.40&75.00&2.33&6.44&1.53&27.60&1.79&106.00&4.62\\
8&1.78&0.30&91.20&9.67&3.36&0.73&17.00&0.86&2.38&0.16&10.40&0.37\\
9&5.35&0.45&323.00&16.20&17.20&0.86&43.30&2.04&6.58&0.56&28.80&1.84\\
10&0.87&0.22&28.60&2.50&2.06&0.20&6.01&0.28&0.53&0.09&2.89&0.06\\
11&4.14&0.22&361.00&9.36&29.60&2.06&32.10&0.95&8.50&0.49&12.60&0.91\\
12&0.72&0.15&78.00&2.50&8.39&0.39&7.90&0.22&0.48&0.33&0.85&0.03\\
13&3.41&0.15&149.00&6.55&5.52&0.20&25.10&0.75&4.33&0.26&17.60&0.51\\
14&9.12&1.50&608.00&62.40&75.60&6.25&58.30&1.53&9.79&0.39&62.30&1.87\\
15&-&-&134.00&15.60&8.31&1.20&11.90&55.40&1.60&0.26&3.71&0.10\\
16&-&-&88.10&3.12&9.39&2.19&5.04&0.13&1.32&0.09&3.16&0.10\\
17&7.70&0.67&170.00&12.80&7.88&1.13&47.30&1.85&11.60&1.03&62.40&2.69\\
18&4.17&0.30&112.00&7.80&10.4&0.73&2.76&0.07&5.30&0.19&24.10&0.27\\
19&2.05&0.30&200.00&12.20&14.90&1.20&29.10&1.25&2.86&0.26&9.55&0.47\\
20&68.80&6.75&985.00&45.20&121.00&4.52&178.00&2.15&85&6.94&289.00&9.55\\
21&-&-&108.00&6.86&11.00&1.26&12.40&0.62&2.31&0.29&18.70&0.64\\
22&7.52&0.45&494.00&22.20&48.70&4.26&68.40&1.45&11.20&1.13&60.30&2.07\\
23&16.90&2.48&636.00&39.90&33.10&2.59&53.70&1.81&11.10&1.10&64.40&4.52\\
24&-&-&113.00&14.40&6.55&0.73&9.21&0.38&0.51&0.06&1.87&0.30\\
25&-&-&15.70&4.06&-&-&3.36&0.21&-&-&-&-\\
26&-&-&60.40&4.37&8.58&0.86&6.21&0.15&0.90&0.09&-&-\\
27&-&-&27.70&4.37&2.33&0.39&3.37&0.23&-&-&1.50&0.23\\
28&1.78&0.52&117.00&12.80&8.39&1.40&11.70&0.69&2.57&0.26&10.80&0.47\\
29&19.40&1.20&1850.00&85.80&271.00&22.70&274.00&10.60&66.50&5.78&321.00&13.00\\
30&77.10&8.55&1900.00&174.00&382.00&39.40&322.00&9.52&125.00&10.30&535.00&15.90\\
31&2.57&0.45&272.00&27.50&25.40&3.46&21.40&1.26&4.94&0.59&27.30&1.56\\
32&27.50&2.55&752.00&39.60&131.00&10.70&84.70&3.78&25.60&1.99&178.00&6.56\\
33&-&-&83.40&7.18&13.70&1.20&9.06&0.31&2.15&0.23&11.10&0.88\\
34&-&-&2.38&1.25&-&-&-&-&-&-&-&-\\
35&-&-&-&-&-&-&0.23&0.05&-&-&-&-\\
36&-&-&8.27&2.18&0.67&0.13&1.28&0.14&-&-&-&-\\
37&5.16&0.90&415.00&31.80&39.40&3.46&49.30&2.21&7.76&0.56&38.50&1.22\\
\hline
\multicolumn{13}{l}{A `-' means that it was not observed emission of such molecule and hence no column density was derived.} \\ 
\end{tabular}
\end{table*}

We measure the integrated flux (S$_{\rm int}$) from a beam towards the peak of each core from the 0.8 mm continuum emission (Col.\,2 in Table\,\ref{sample}).  Using these values we
derive the H$_{2}$ column density (N(H$_{2}$)) of each core to be used to estimate molecular abundances. To carry on this, we use the following equation \citep{kauff08}:

\begin{eqnarray}
{\rm N(H_{2})=2.02 \times 10^{20}~cm^{-2} \left(e^{1.439(\lambda/mm)^{-1}(T/10~K)^{-1}} - 1 \right) } \nonumber \\ 
{\rm \times \left( \frac{\lambda}{mm}\right)^{3}\left(\frac{\kappa_{\nu}}{0.01 ~cm^{2}g^{-1}} \right)^{-1}\left(\frac{S_{int} }{mJy} \right) \left(  \frac{\theta_{HPBW}}{10~arcsec} \right)    }
\label{NH2}
\end{eqnarray}

\noindent where T is the dust temperature and $\kappa_{\nu}$ is the dust opacity per gram of matter at 870~$\mu$m, for which we adopt the value of 0.0185~cm$^2$g$^{-1}$ \citep[][and references therein]{csengeri2017}, and $\theta_{\rm HPBW}$ is the beam size. We assume thermal coupling between dust and gas (i.e. T$_{\rm dust}$=T$_{\rm kin}$). To obtain a value of T$_{\rm kin}$ for each core we look for other molecular lines in the data set useful to estimate temperatures. We found that all cores have well-defined emission of the CH$_{3}$OH 7(1,7)--6(1, 6)++ and 12(1,11)--12(0,12)-+ lines at 335.582 and 336.865 GHz (for an example of spectra of these lines, see Fig.\,\ref{C4} in Appendix\,\ref{example}). Then, by integrating the emission of such lines (presented in Table\,\ref{metanol}), from a typical rotational diagram procedure (\citealt{goldsmith99}; see Appendix\,\ref{appmetanol}) we obtain the rotational temperatures that were assumed to be the T$_{\rm kin}$ of each core (Col.\,4, and their errors in Col.\,5 of Table\,\ref{sample}). The N(H$_{2}$) obtained for each core with the errors are presented in Cols.\,6, and 7 of Table\,\ref{sample}. 

Additionally, to have a rough characterization of the cores, we calculated the mass using the continuum fluxes measured from a beam towards the peaks of the cores. The mass was estimated from \citep{kauff08}:

\begin{eqnarray}
{\rm M_{gas} = 0.12~M_{\odot}\left[exp\left(\frac{1.439}{(\lambda/mm)(T/10~K)}\right) -1 \right]} \nonumber \\
{\rm \times  \left(\frac{\kappa_{\nu}}{0.01 ~cm^{2}g^{-1}} \right)^{-1}\left(\frac{S_{int} }{mJy} \right) \left( \frac{d}{100~pc} \right)^{2} \left( \frac{\lambda}{mm}\right)^{3}   }
\label{mass}
\end{eqnarray}

The volume densities were derived assuming spherical symmetries with a diameter of the beam (i.e. the density was obtained for a uniform probe volume at the peak of each core). Masses and densities are included in the last columns of Table\,\ref{sample}. In these cases, errors are not quoted given that not all cores have errors in their distances, and these parameters are taken as a rough core characterization in this work.

\begin{table*}[h]
\centering
\caption{Measured and derived core parameters.}
\label{sample}
\begin{tabular}{lcccccccc}
\hline
\hline
Core &S$_{int}$& Error & T$_{\rm CH_{3}OH}$& Error & N(H$_{2}$)& Error & Mass$^{\dagger}$ &n(H$_{2}$)$^{\dagger}$ \\
&(Jy)& &(K)&  & ($\times 10^{23}$cm$^{-2}$)& & (M$_{\odot}$)&($\times 10^{6}$cm$^{-3}$) \\
\hline
1&1.91&0.18&74.1&2.1&1.17&0.14&-&-\\
2&6.24&0.24&63.7&2.1&4.55&0.34&155.6&9.82\\
3&1.59&0.06&79.4&3.8&0.90&0.08&31.0&1.96\\
4&0.93&0.06&86.2&0.9&0.48&0.03&11.7&1.24\\
5&1.58&0.05&27.1&3.1&3.33&0.62&30.5&13.50\\
6&1.41&0.08&105.2&4.3&0.58&0.05&22.1&1.22\\
7&4.44&0.15&77.5&1.9&2.59&0.15&26.4&10.30\\
8&2.04&0.07&69.9&3.2&1.34&0.11&12.5&5.53\\
9&6.80&0.25&60.9&1.2&5.23&0.30&11.2&45.00\\
10&1.47&0.05&61.7&2.0&1.11&0.07&9.5&4.80\\
11&3.06&0.06&66.2&0.9&2.14&0.07&83.4&4.31\\
12&2.16&0.11&49.2&0.8&2.13&0.14&75.9&4.48\\
13&3.80&0.09&60.9&1.1&2.92&0.12&56.1&8.39\\
14&3.88&0.07&66.2&1.9&2.71&0.13&101.3&5.59\\
15&2.56&0.11&81.9&1.2&1.41&0.08&50.4&2.97\\
16&1.58&0.03&74.6&2.0&0.96&0.04&40.9&1.87\\
17&1.99&0.05&86.2&1.6&1.03&0.04&10.5&4.10\\
18&1.63&0.07&69.9&1.0&1.07&0.06&115.0&1.30\\
19&2.30&0.12&49.2&0.7&2.27&0.15&63.9&5.36\\
20&6.81&0.19&76.9&0.9&4.01&0.16&113.6&9.54\\
21&2.41&0.08&69.4&1.7&1.59&0.09&45.1&3.79\\
22&2.58&0.10&72.9&2.3&1.61&0.11&83.3&2.84\\
23&10.04&0.47&55.2&0.9&8.65&0.57&111.0&30.30\\
24&3.10&0.16&43.8&1.0&3.52&0.27&55.6&11.10\\
25&1.02&0.04&47.3&1.5&1.06&0.08&9.8&4.33\\
26&1.06&0.09&76.3&3.6&0.63&0.08&9.4&2.07\\
27&3.19&0.18&40.0&0.8&4.05&0.33&68.2&12.30\\
28&1.92&0.08&77.5&4.1&1.12&0.11&15.6&3.81\\
29&6.90&0.27&80.0&1.8&3.89&0.25&66.2&11.90\\
30&12.38&0.78&84.0&4.5&6.61&0.80&105.4&21.00\\
31&3.82&0.30&68.9&3.3&2.55&0.33&79.4&5.76\\
32&9.16&0.38&90.9&3.4&4.48&0.37&62.3&15.20\\
33&1.18&0.16&88.4&4.1&0.59&0.11&4.2&2.84\\
34&1.07&0.08&21.4&0.4&3.15&0.32&9.1&22.30\\
35&1.73&0.09&-&-&-&-&-&-\\
36&1.90&0.08&44.0&2.0&2.15&0.21&6.4&15.60\\
37&4.45&0.32&59.5&2.0&3.51&0.38&367.3&4.32\\
\hline
 & & &  & & &  \\[-2ex]
\multicolumn{9}{l}{$^{\dagger}$ Errors in mass and volume density are not quoted (see text).} \\ 
\end{tabular}
\end{table*}

\subsection{Molecular abundances}

By defining the abundance of a molecular species as X(molecule) = N(molecule)/N(H$_{2}$), we calculated such values and performed comparisons among them and the physical parameters for each core.

\begin{figure*}
    \centering
   \includegraphics[width=6cm]{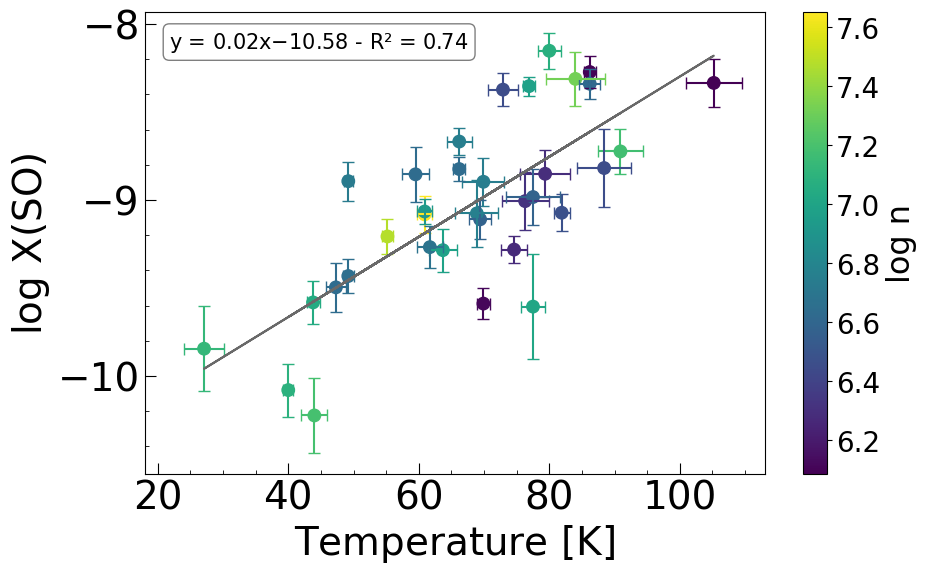}
    \includegraphics[width=6cm]{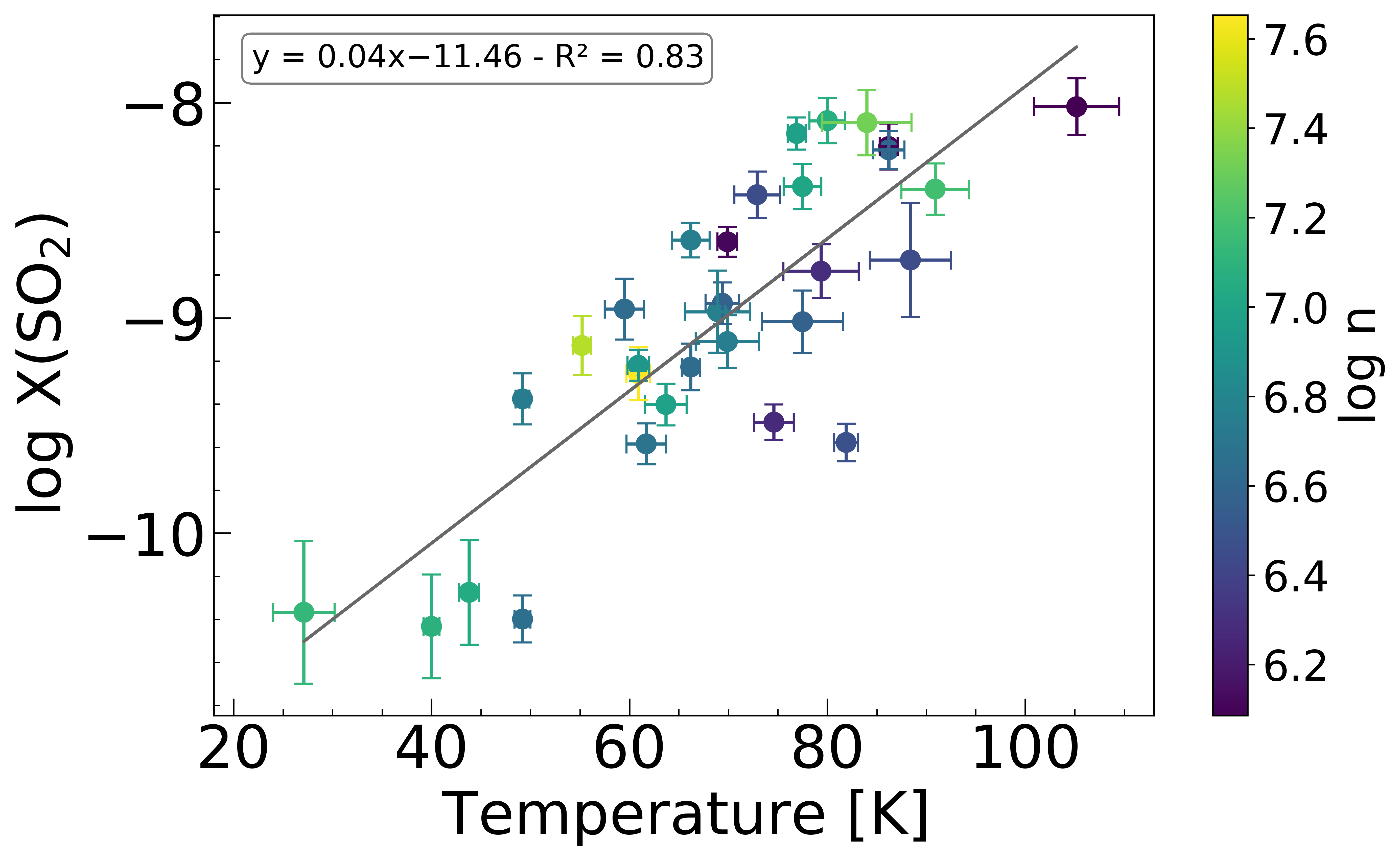}
    \includegraphics[width=6cm]{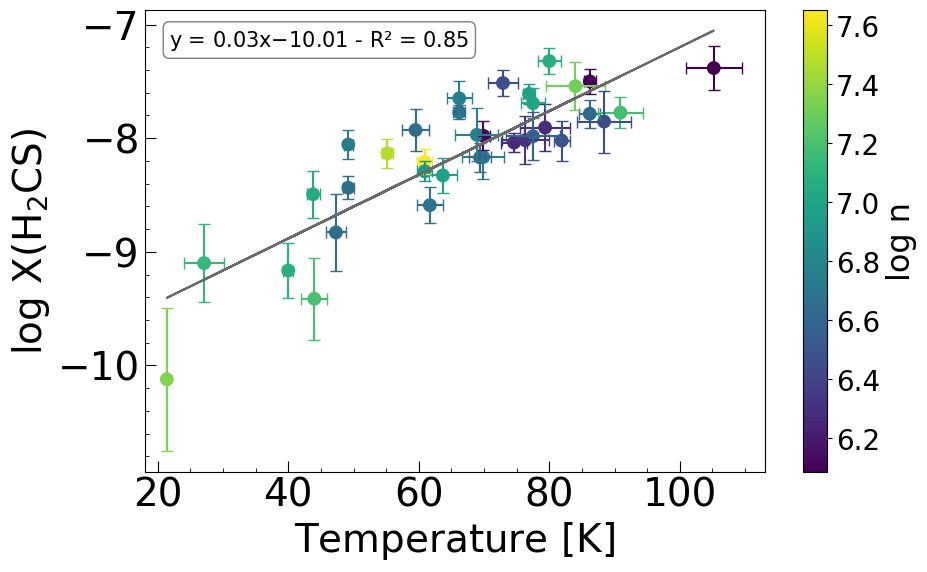}
    \includegraphics[width=6cm]{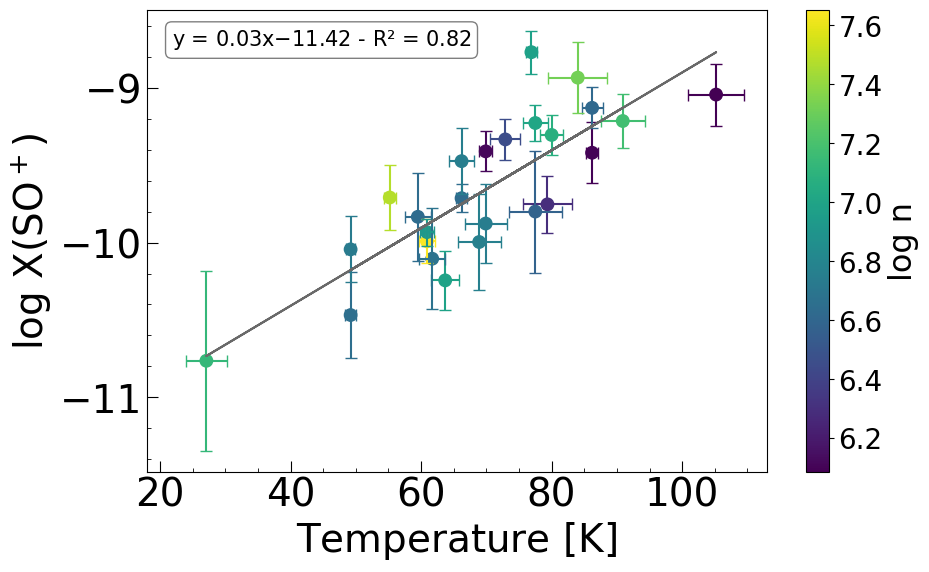}
    \includegraphics[width=6cm]{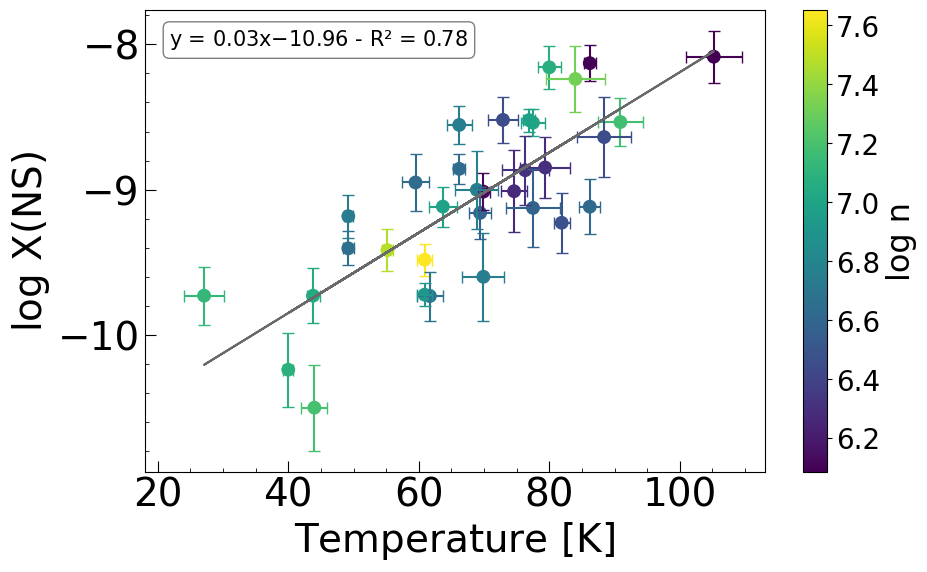}
    \includegraphics[width=6cm]{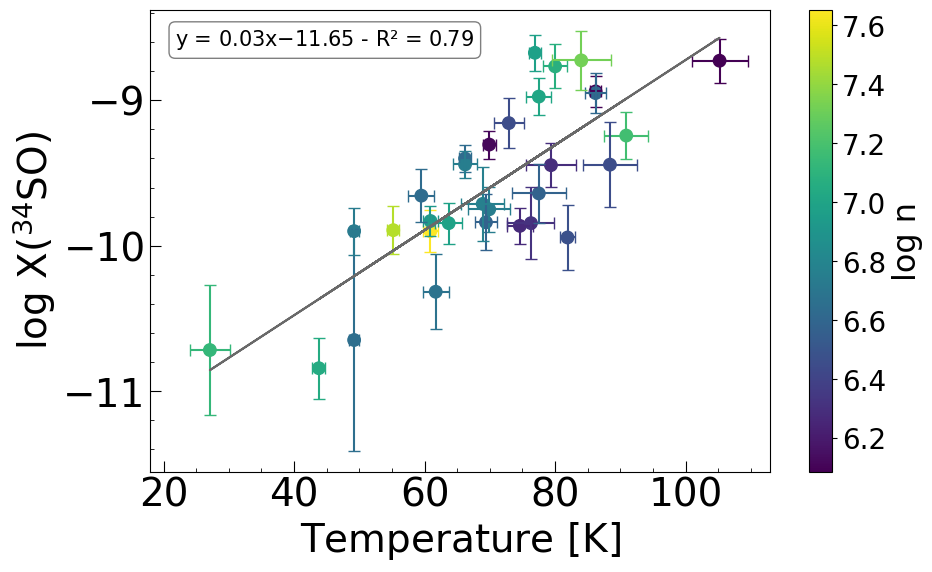}
        \caption{Molecular abundances (X(x) = N(x)/N(H$_{2}$)) (in logarithmic scale) vs. kinetic temperature. The lines are the result of linear fittings whose results are included at the top left corner of each panel. The colors correspond to the volume density of the cores (in cm$^{-3}$), and the scale is presented in the bar at the right of each panel. }
    \label{XvsT}
\end{figure*}

\begin{figure*}
    \centering
    \includegraphics[width=6cm]{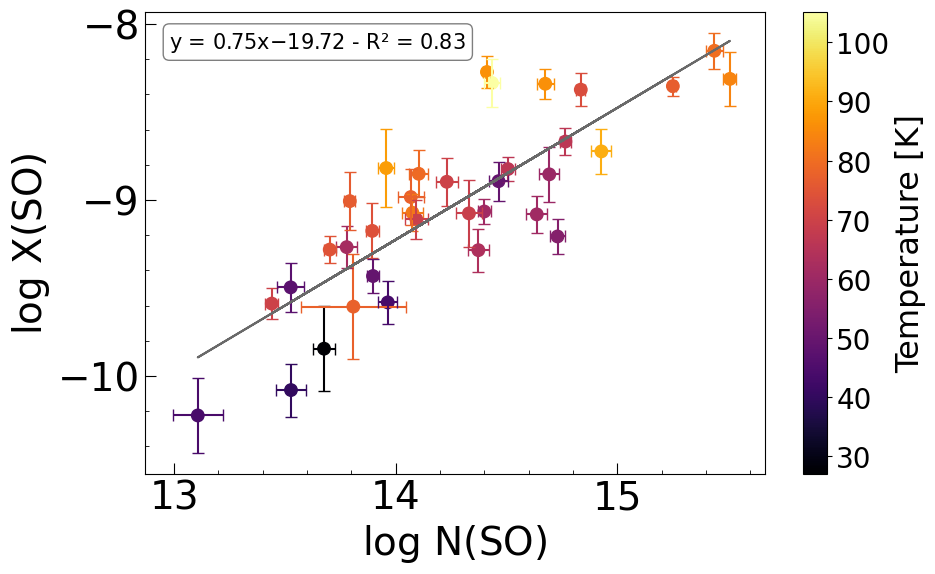}
   \includegraphics[width=6cm]{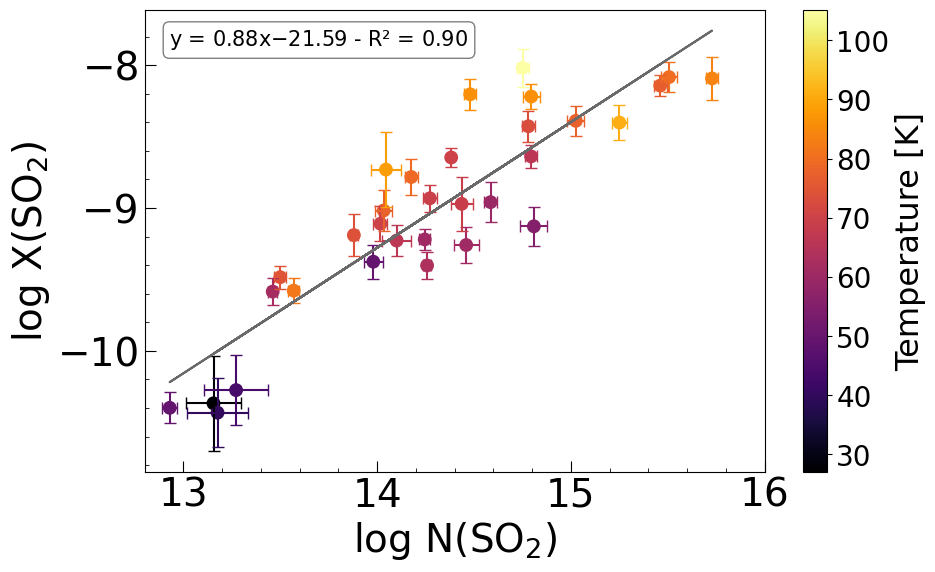}
   \includegraphics[width=6cm]{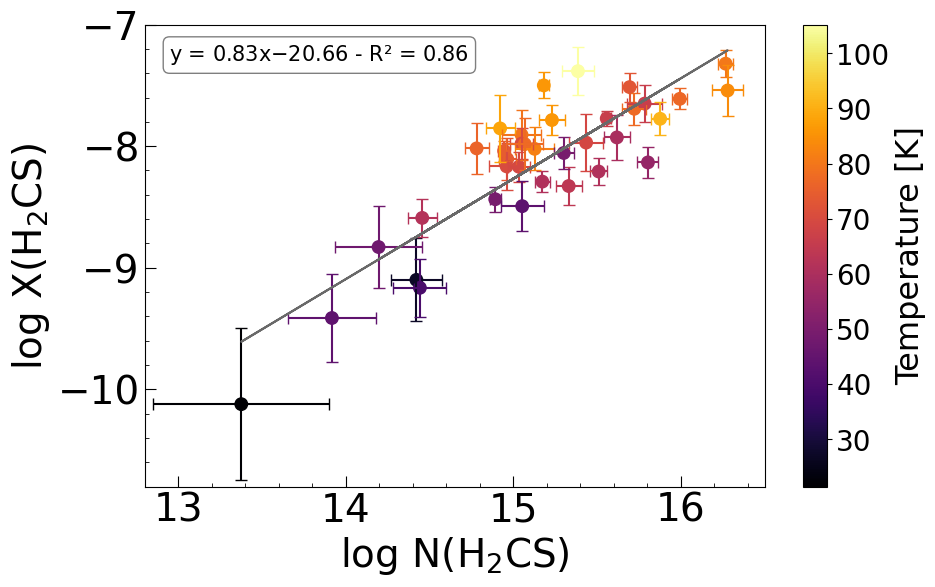}
   \includegraphics[width=6cm]{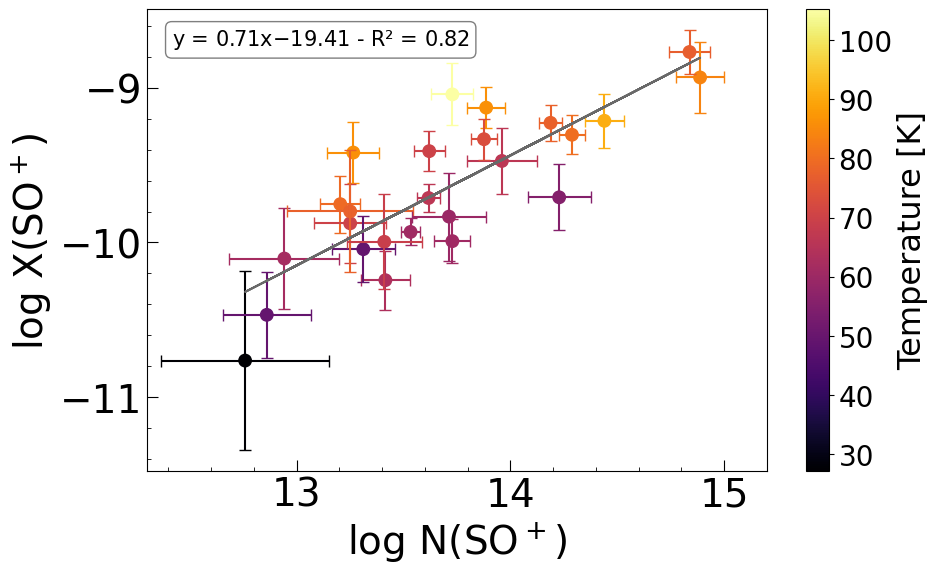}
   \includegraphics[width=6cm]{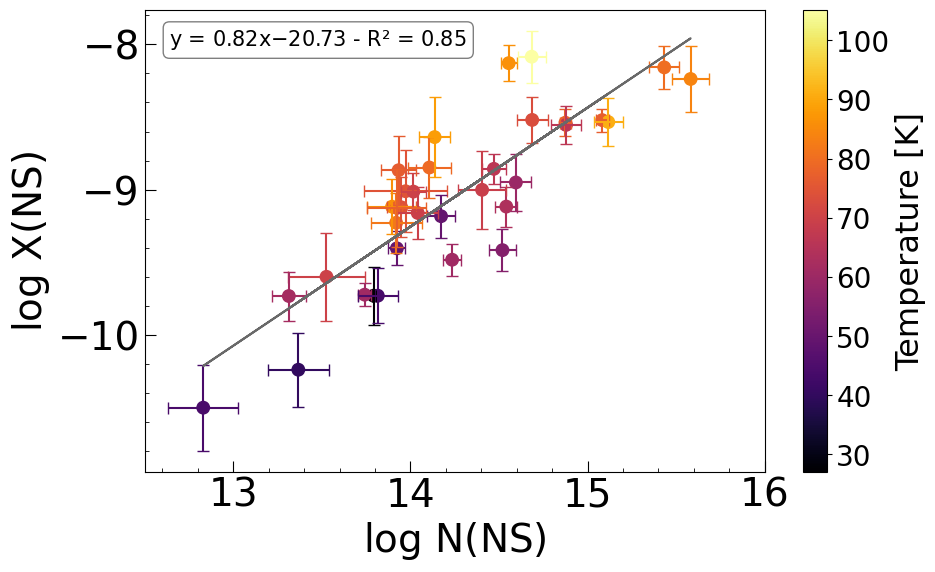}
   \includegraphics[width=6cm]{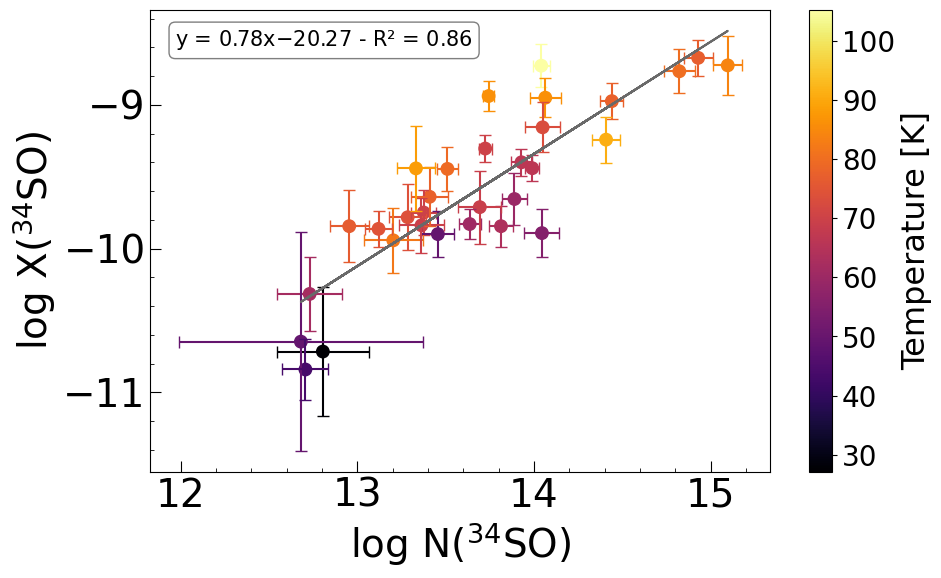}
    \caption{Molecular abundances  (X(x) = N(x)/N(H$_{2}$)) vs. column density (N(x)) presented in logarithmic scale. The lines are the result of linear fittings whose results are included at the top left corner of each panel. The colors correspond to the kinetic temperature measured in the cores, and the scale is presented in the bar at the right of each panel.   }
    \label{XvsNx}
\end{figure*}

\begin{figure*}
    \centering
    \includegraphics[width=6cm]{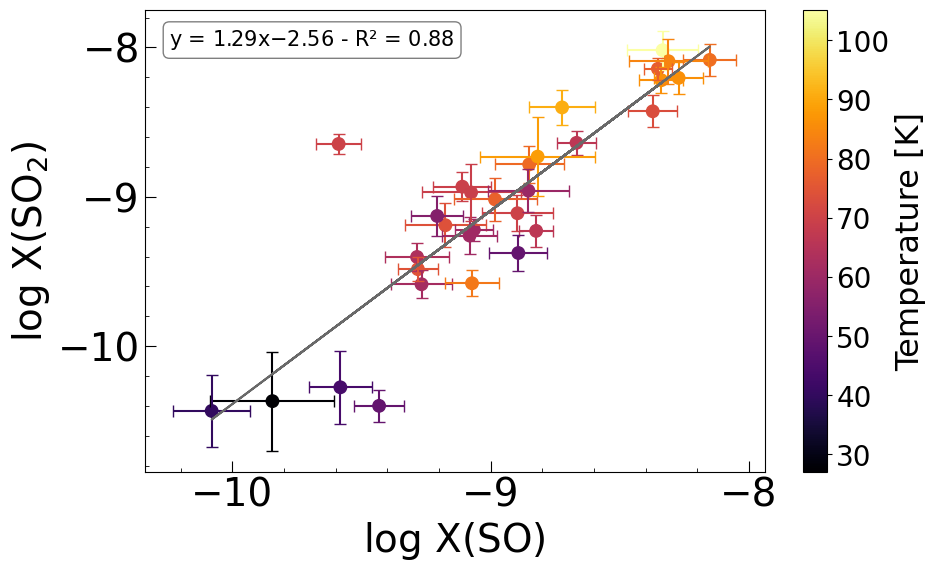}
    \includegraphics[width=6cm]{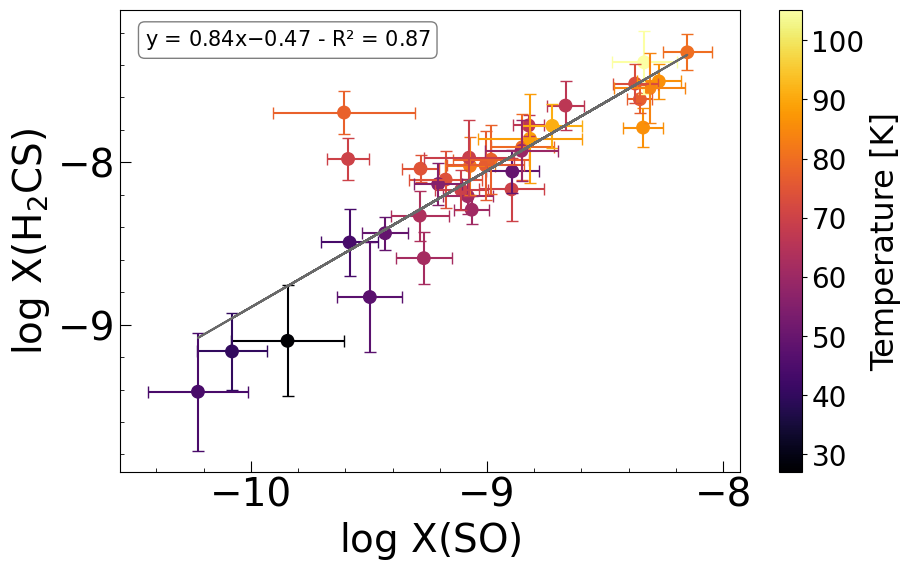}
    \includegraphics[width=6cm]{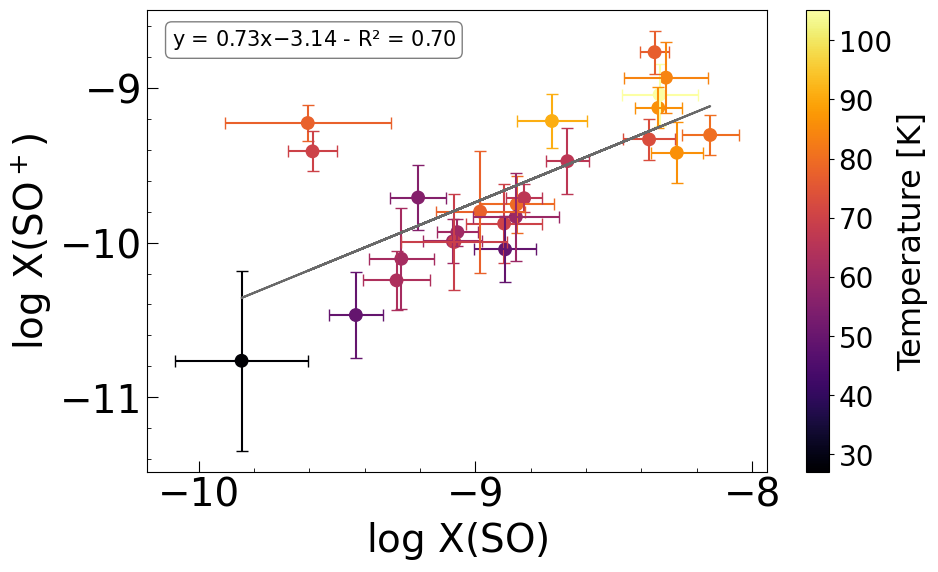}
    \includegraphics[width=6cm]{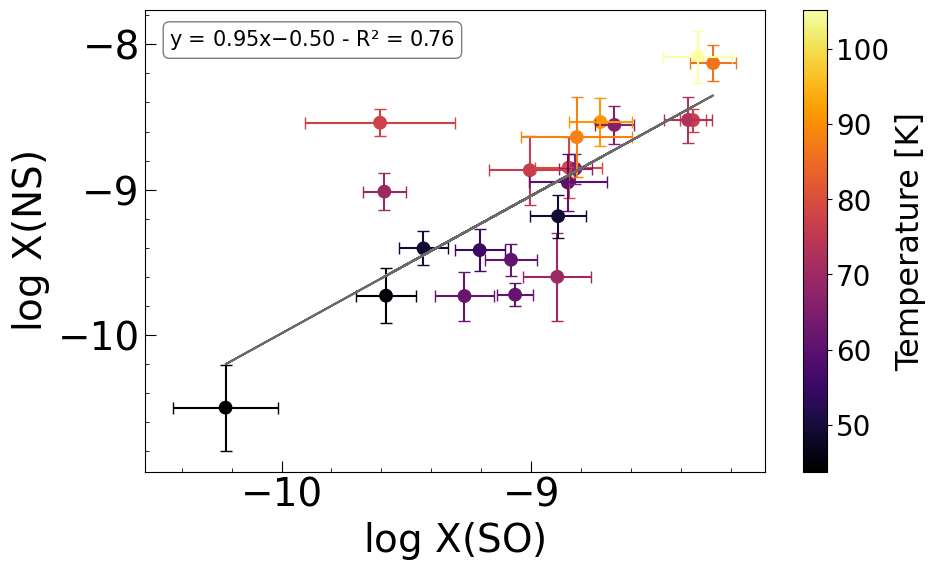}
    \includegraphics[width=6cm]{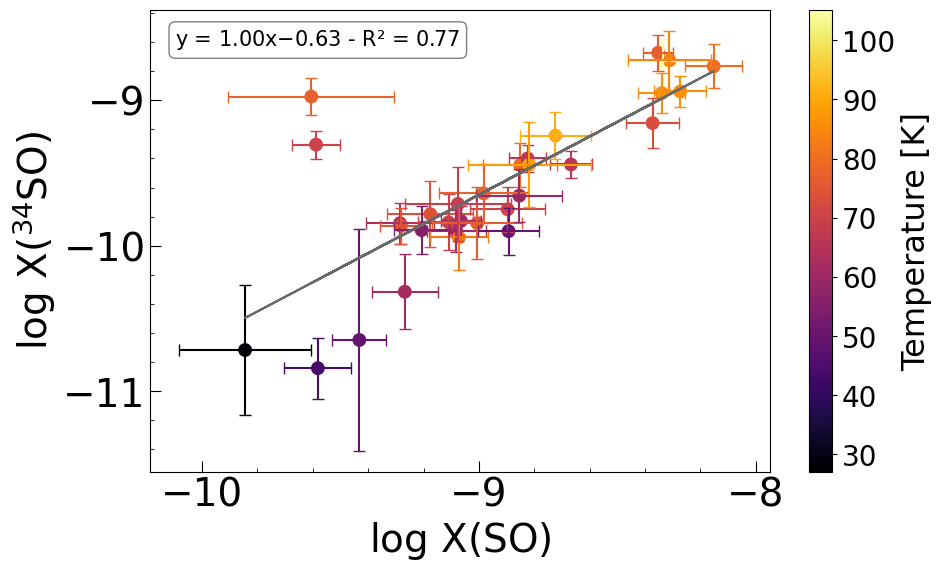}
    \includegraphics[width=6cm]{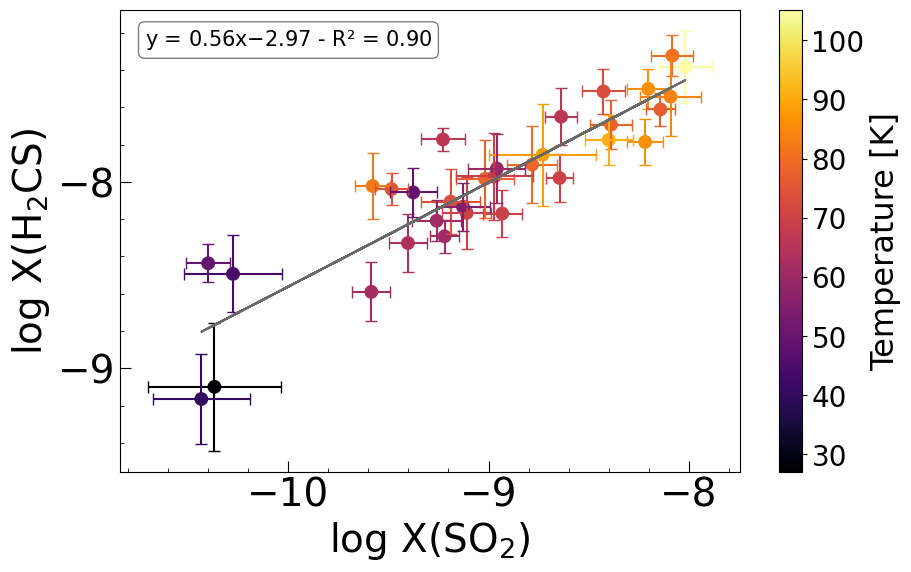}
    \includegraphics[width=6cm]{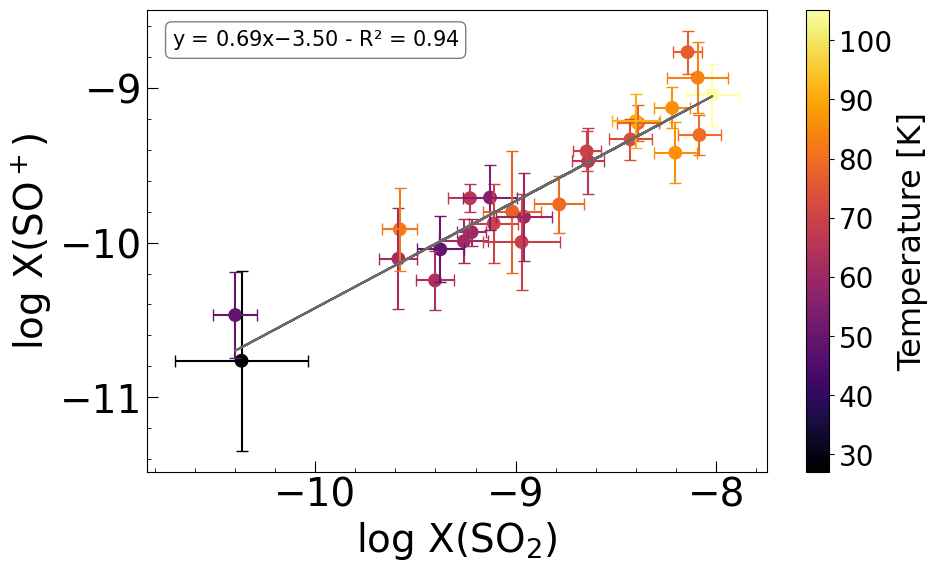}
    \includegraphics[width=6cm]{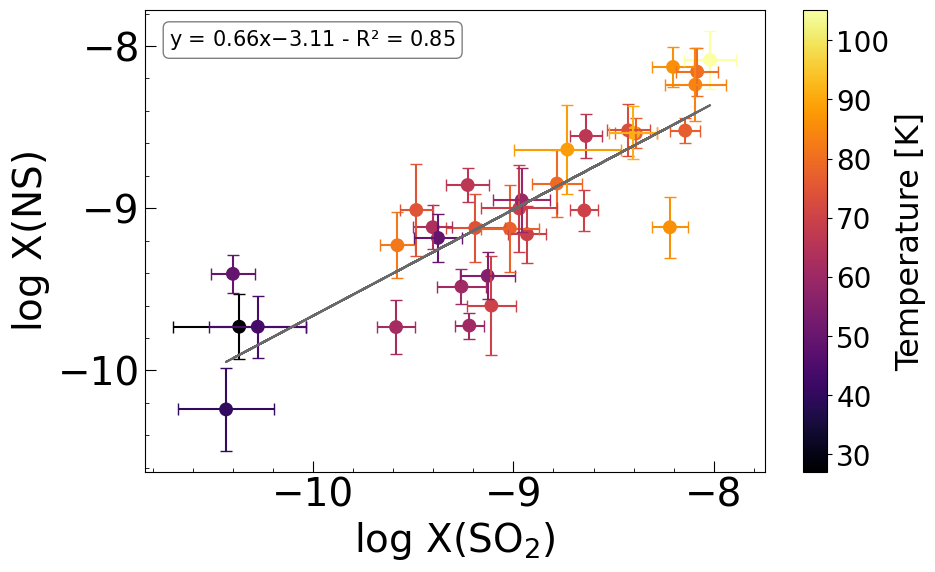}
    \includegraphics[width=6cm]{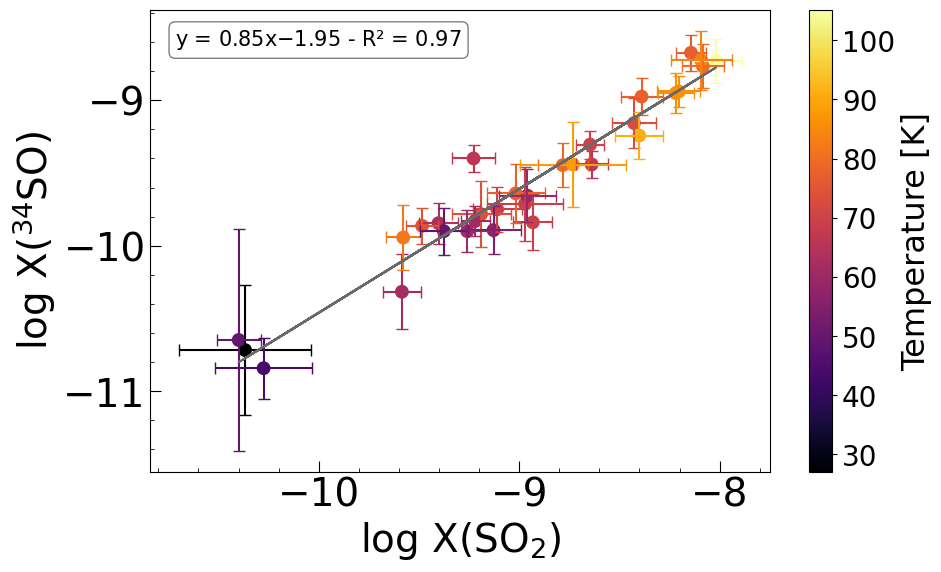}
    \includegraphics[width=6cm]{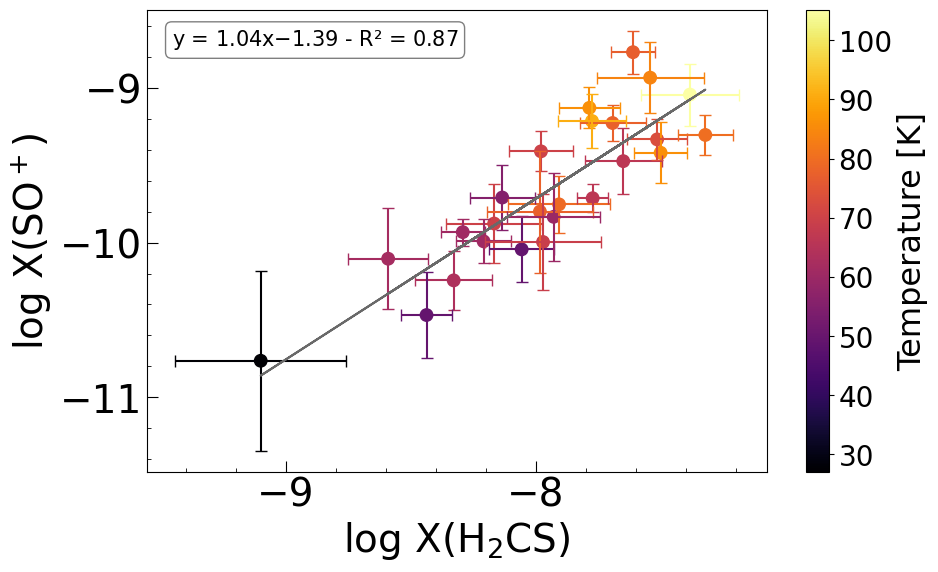}
    \includegraphics[width=6cm]{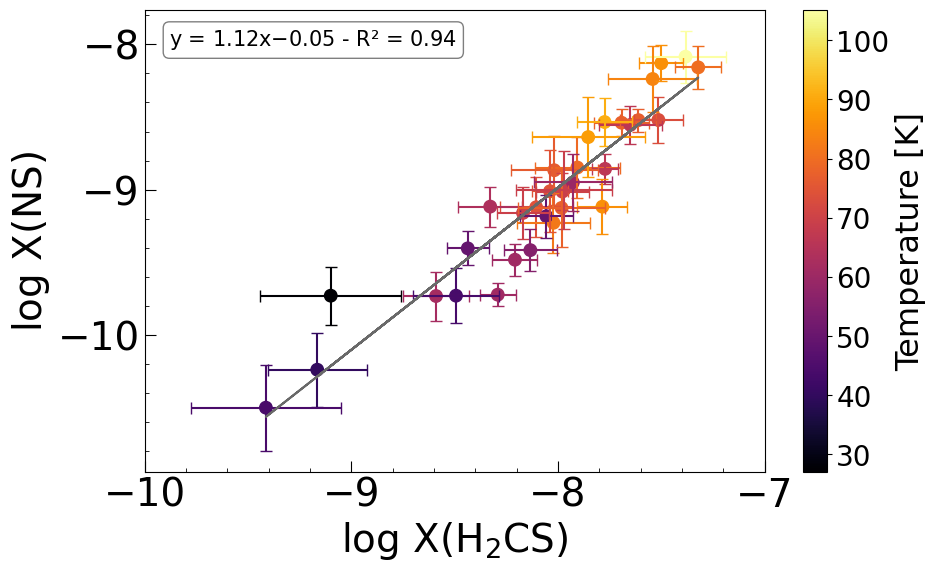}
    \includegraphics[width=6cm]{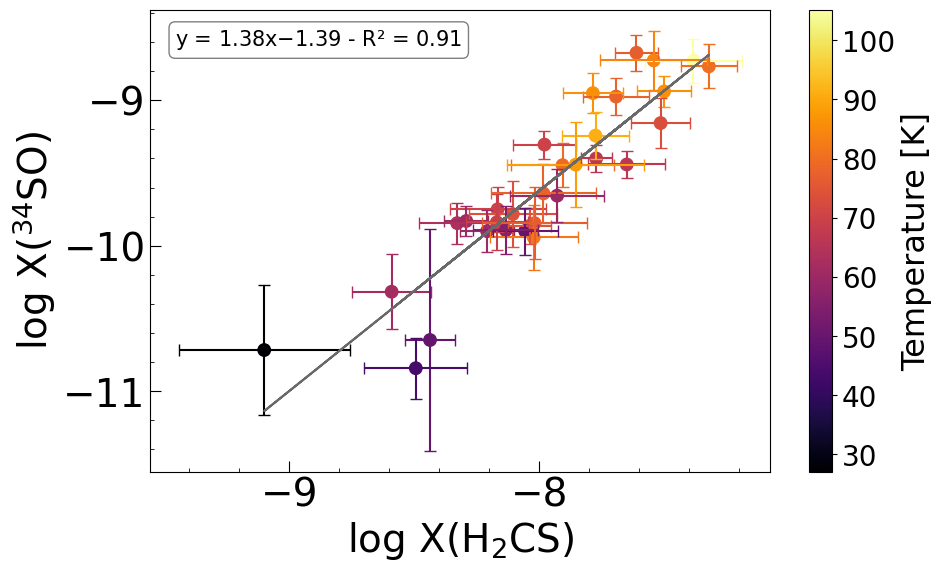}
    \includegraphics[width=6cm]{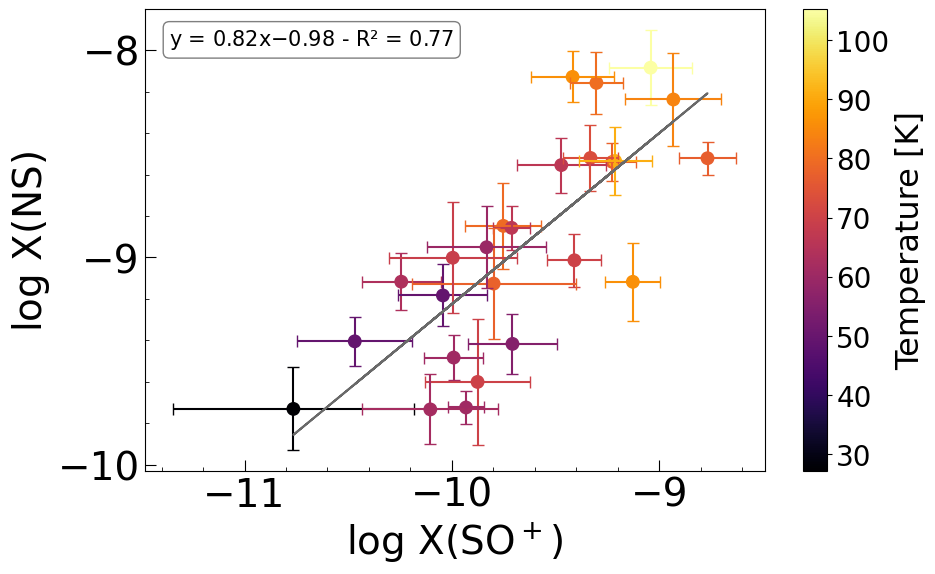}
    \includegraphics[width=6cm]{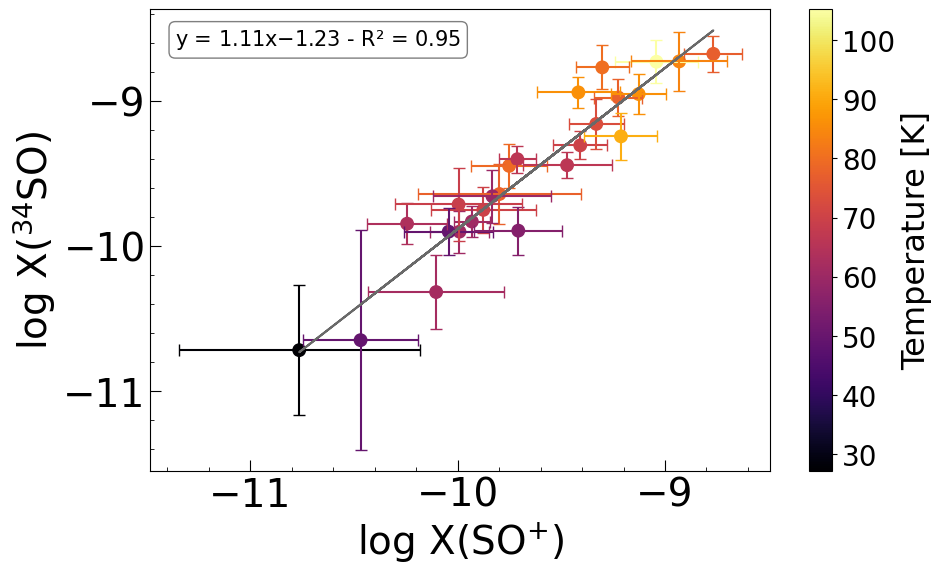}
    \includegraphics[width=6cm]{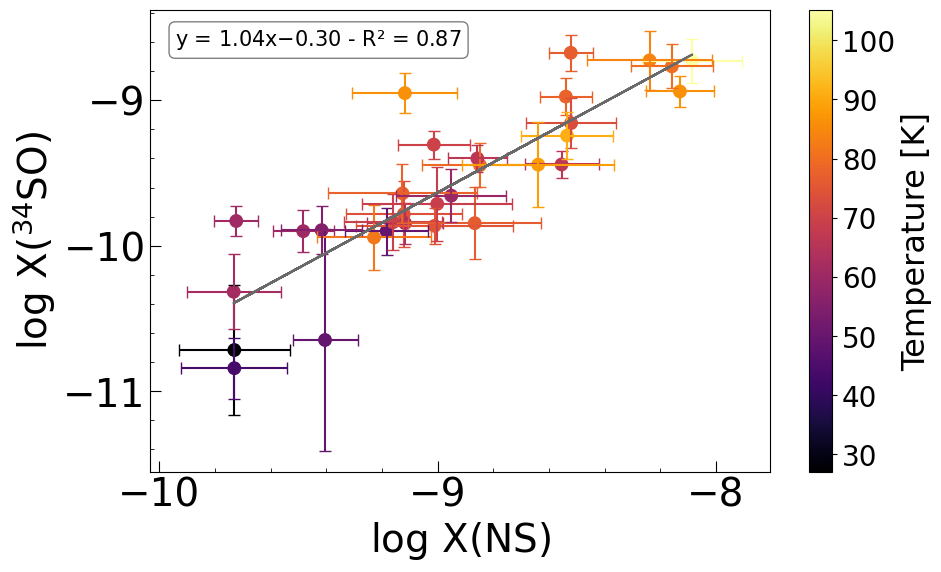}
    \caption{Molecular abundances comparisons among all analyzed molecules presented in logarithmic scale. The lines are the result of linear fittings whose results are included at the top left corner of each panel. The colors correspond to the kinetic temperature measured in the cores, and the scale is presented in the bar at the right of each panel. }
    \label{XvsX}
\end{figure*}

Given that the temperature can have deep implications for the chemistry (e.g. \citealt{gorai20}), in Fig.\,\ref{XvsT} we present abundances vs. temperature for each core, putting in a color scale their relations with the density. For its part, Fig.\,\ref{XvsNx} displays X(molecule) vs. N(molecule) relations with a color scale representing temperature. Finally,
Fig.\,\ref{XvsX} shows the relation among all the sulfur-bearing molecular abundances with the color scale representing temperature. All figures present the results of a linear fitting with the coefficient of determination R$^{2}$. 

Given that it was proposed that the abundance ratio X(SO$_{2}$)/X(SO) could be used as a chemical clock (e.g. \citealt{wak04,wak11}), we calculate such a ratio for our sample of cores. We obtain a media of 1.6, with a maximum value of 9.29 and a minimum of 0.06.

\subsection{Molecular line widths}

The molecular line widths $\Delta$v (FWHM) are usually associated with the kinematics of the molecular gas which can reflect the consequences induced by multiple events, such as shocks, outflows, rotation, and turbulence. If we assume that line widths in our sample of cores primarily result from turbulence related to early star formation (e.g. \citealt{sanhueza12,naiping21}), comparing this parameter across different molecular species could provide valuable physical and chemical insights. For instance, it could give information about the location or distribution of the molecular species in different gaseous layers within the core. Indeed, as done and pointed out by \citet{fontani23}, these comparisons may allow us to understand
the species that are more likely associated with similar gas, and to search for variations among the molecular lines. 

\begin{figure*}
    \centering
        \includegraphics[width=0.8\textwidth]{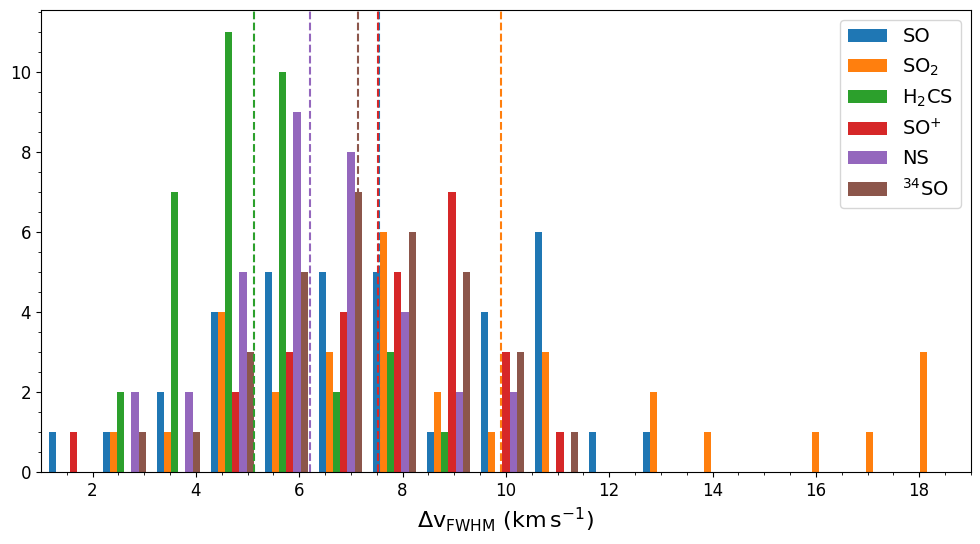}
        \caption{Distribution of the measured line widths $\Delta$v (FWHM) of the sample. The vertical dashed lines are the average values of $\Delta$v of each molecule. In the case of SO and SO$^{+}$, such averages are almost the same, the dashed lines appear overlapped. The horizontal axis is truncated at $20~\mathrm{km\,s^{-1}}$ for a better visualization. }
    \label{histo_DF}
\end{figure*}

In Fig.\,\ref{histo_DF}, we show the velocity width distribution (FWHM) of the six molecular species for the full sample of sources. We identify H$_2$CS as the molecule having the narrower line widths (the distribution peaks at $\sim 5$ \ks), so we can use it as a tracer of the quiescent gas component. On the other hand, SO$_2$ displays the wider widths, indicating that it is tracing the most turbulent gas.    

\begin{figure*}
    \centering
    \includegraphics[width=6cm]{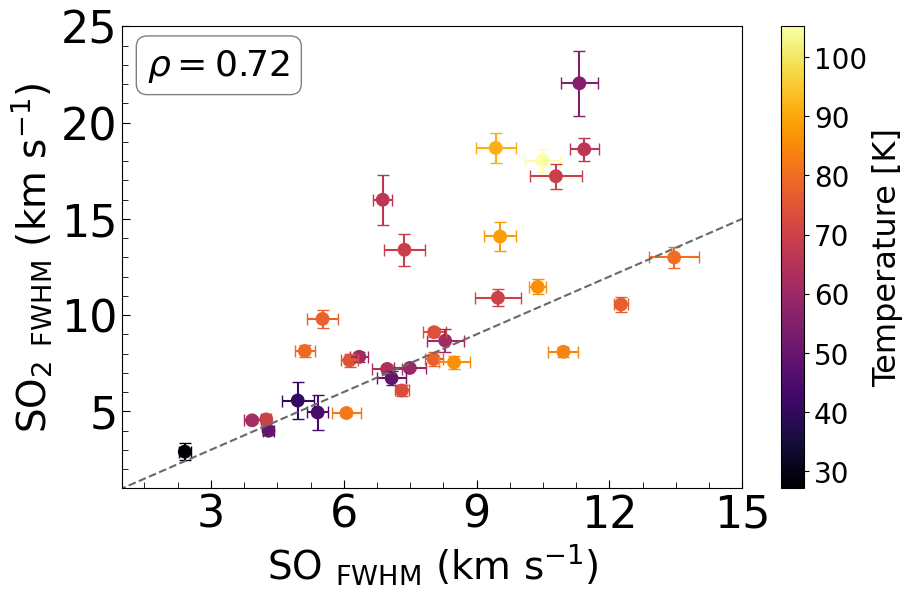}
    \includegraphics[width=6cm]{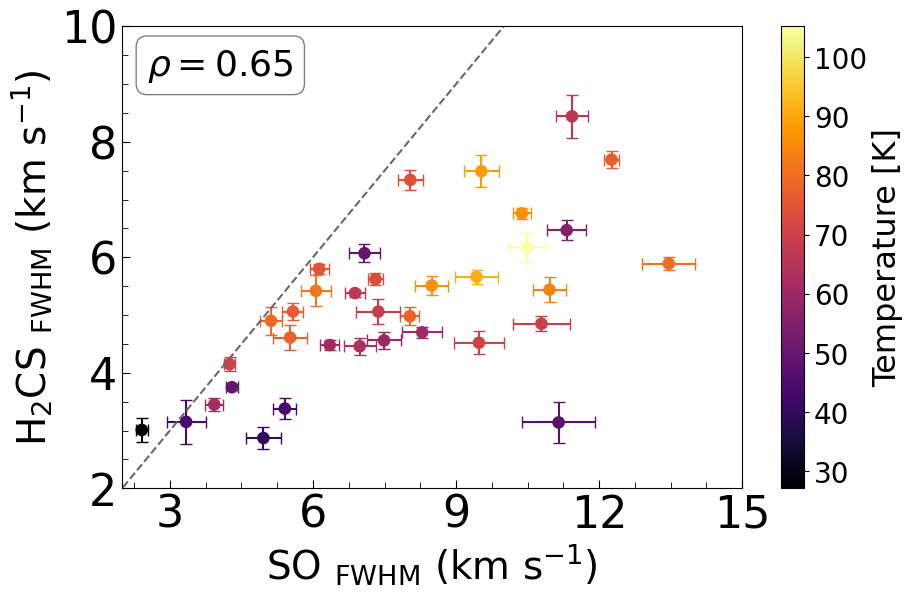}
    \includegraphics[width=6cm]{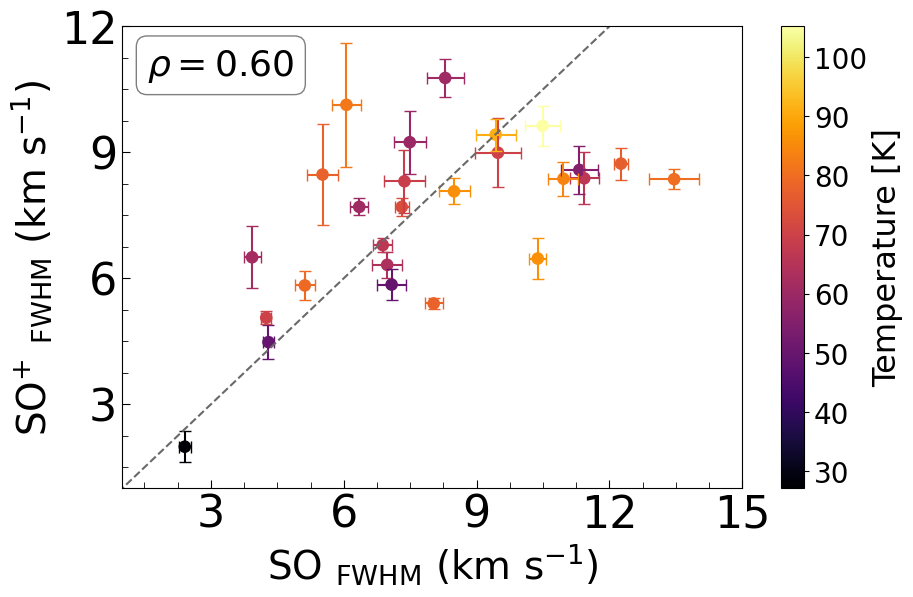}
    \includegraphics[width=6cm]{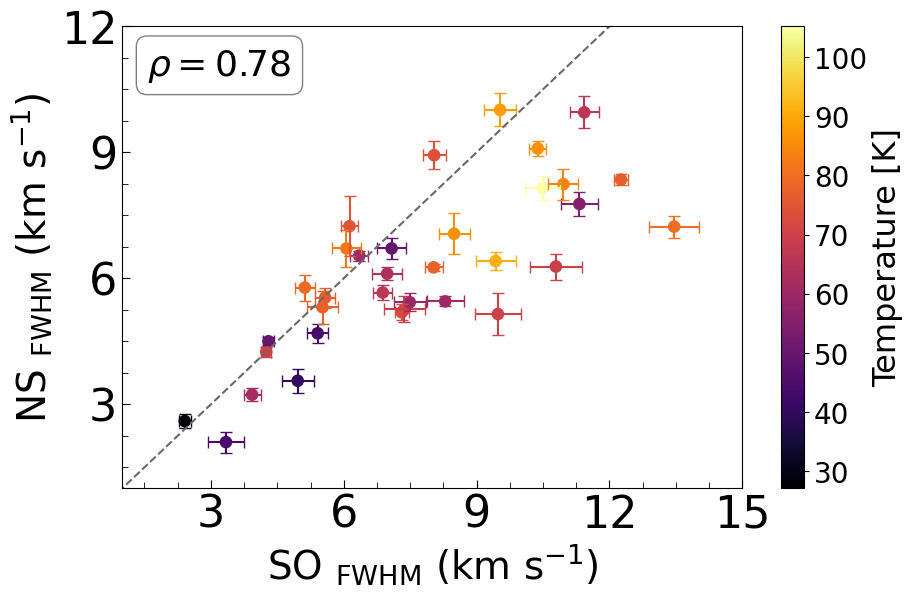}
    \includegraphics[width=6cm]{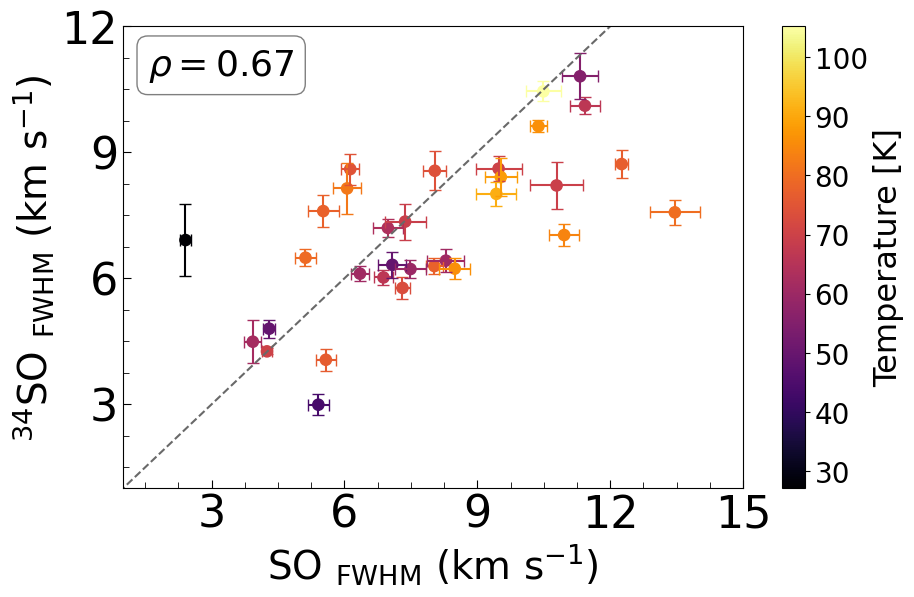}
    \includegraphics[width=6cm]{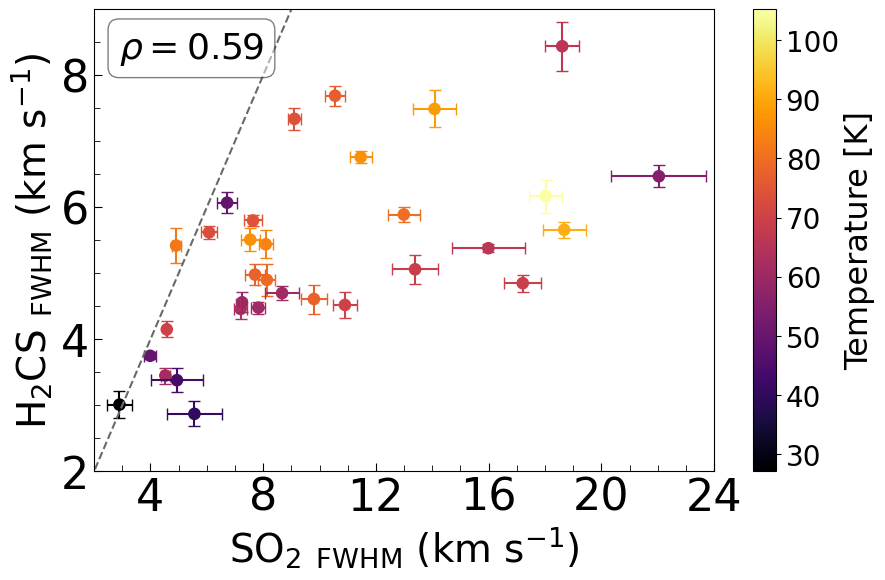}
    \includegraphics[width=6cm]{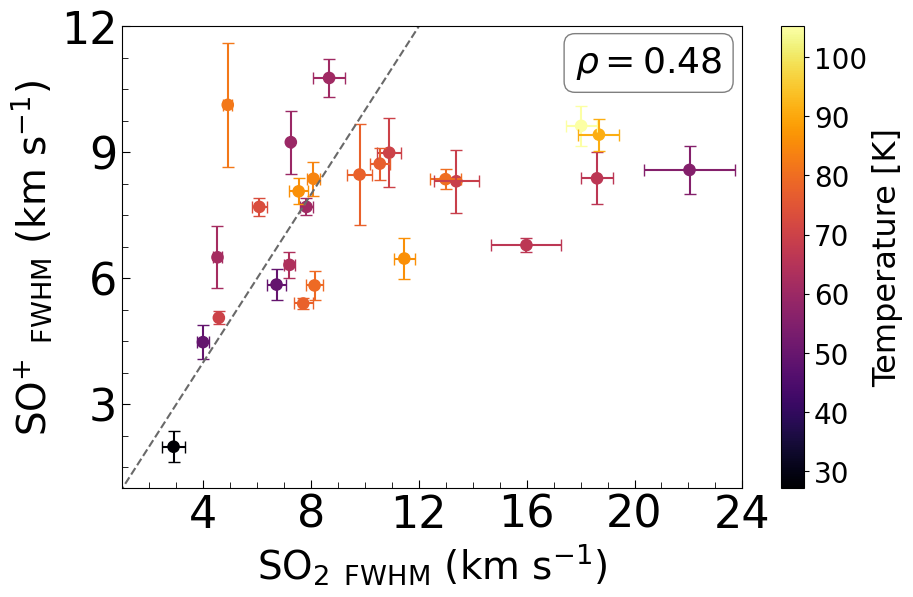}
    \includegraphics[width=6cm]{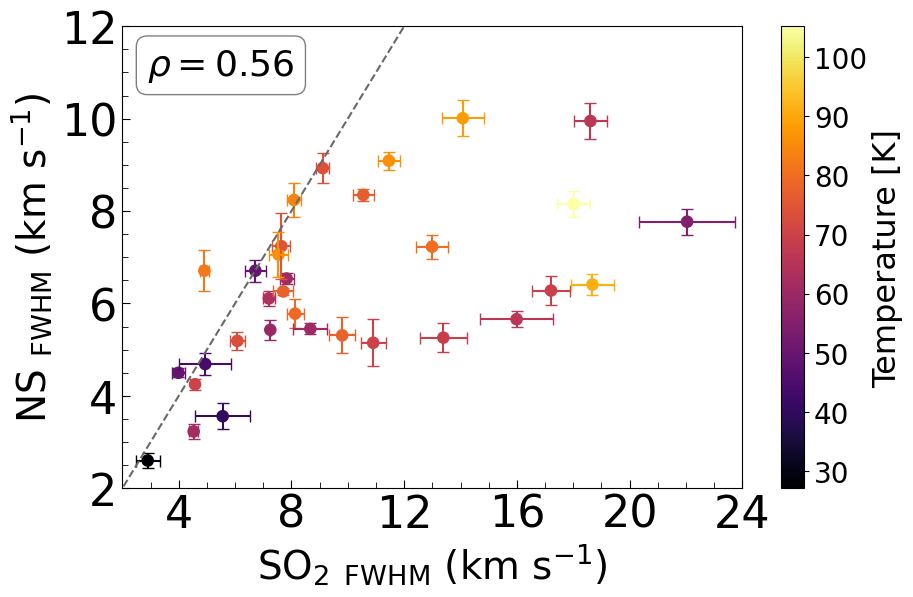}
    \includegraphics[width=6cm]{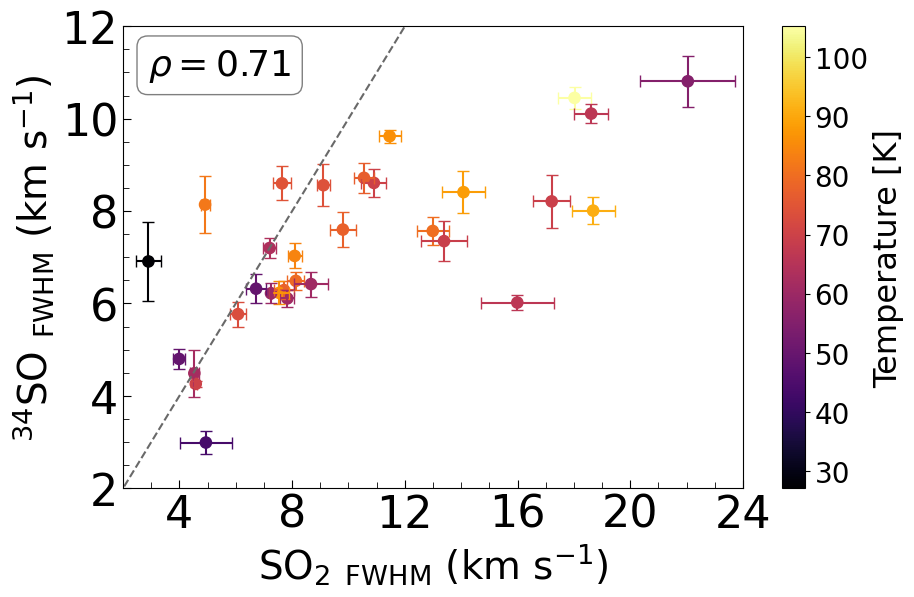}
    \includegraphics[width=6cm]{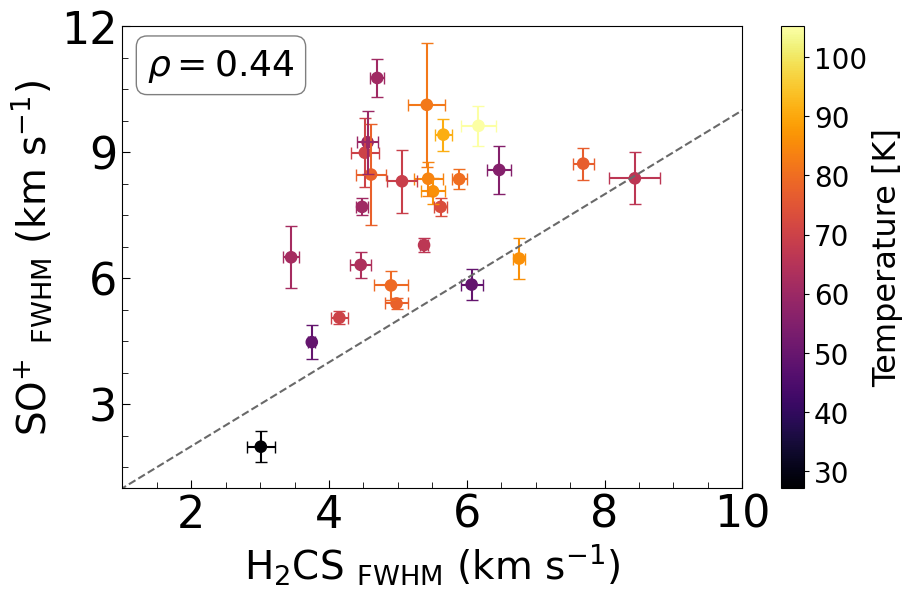}
    \includegraphics[width=6cm]{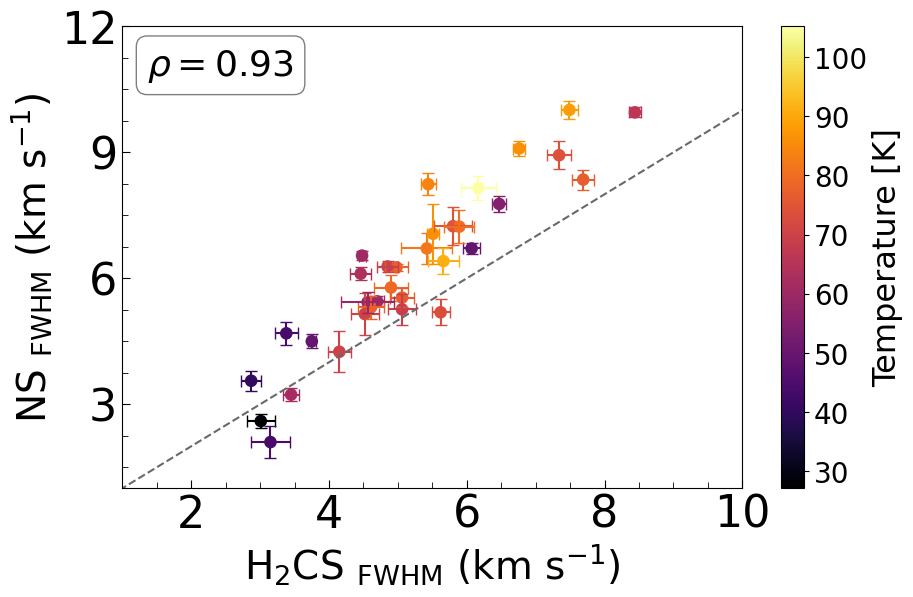}
    \includegraphics[width=6cm]{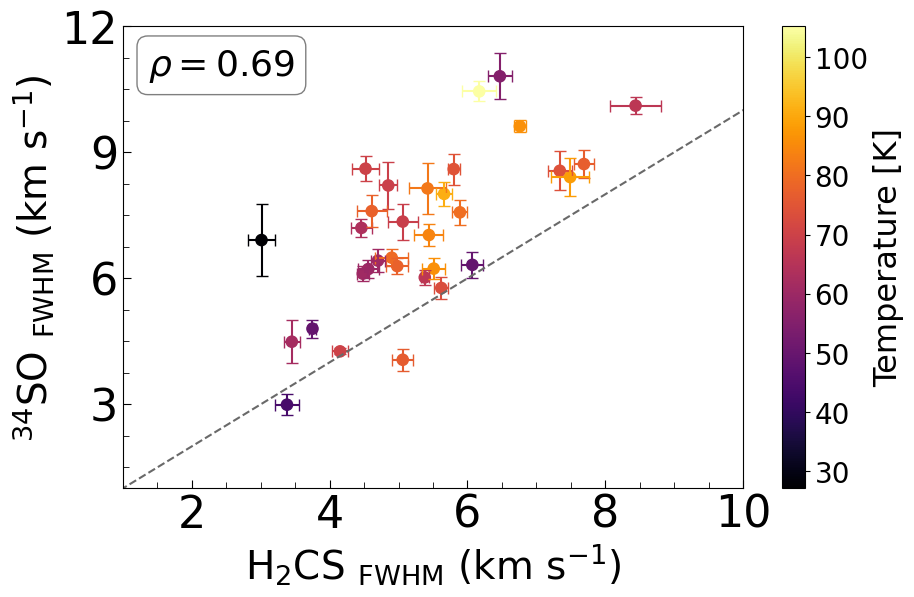}
    \includegraphics[width=6cm]{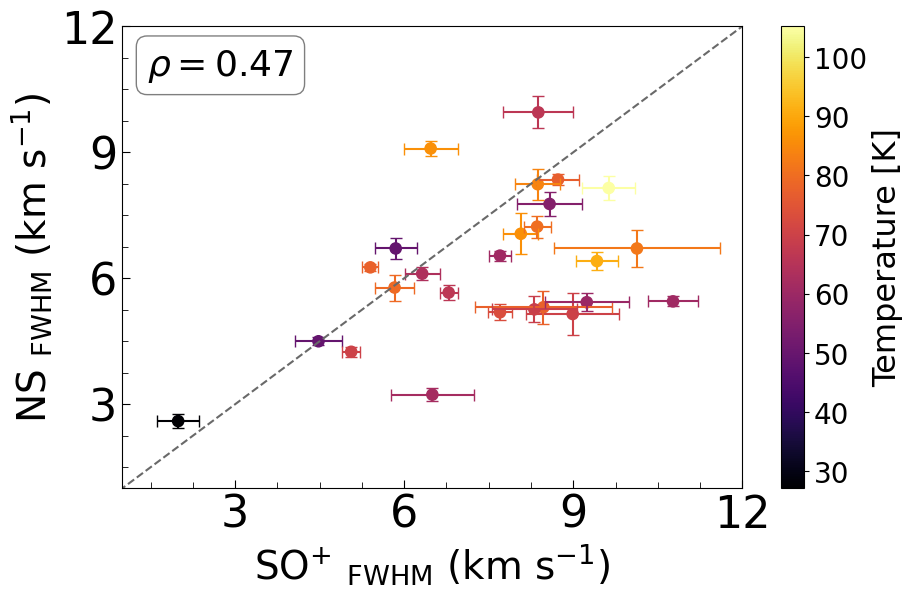}
    \includegraphics[width=6cm]{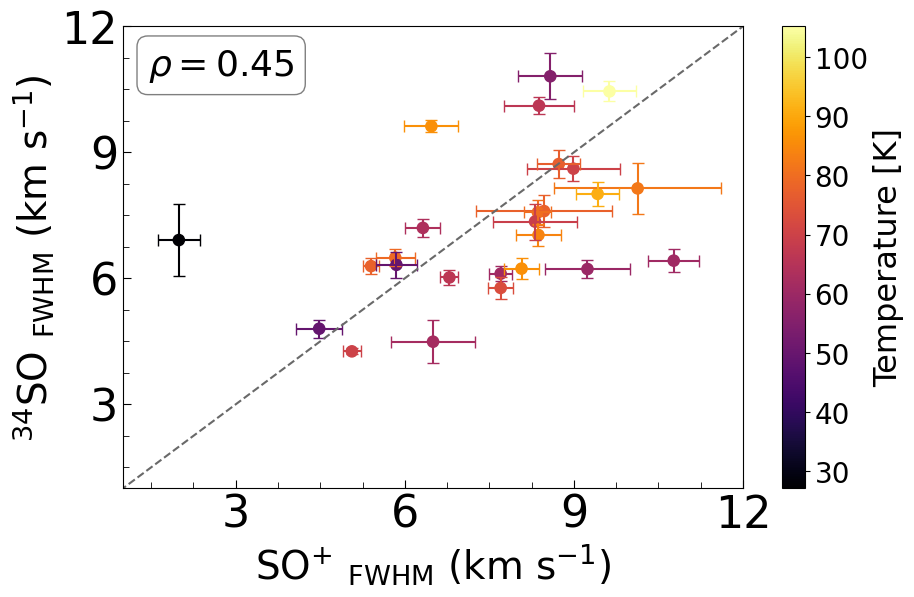}
    \includegraphics[width=6cm]{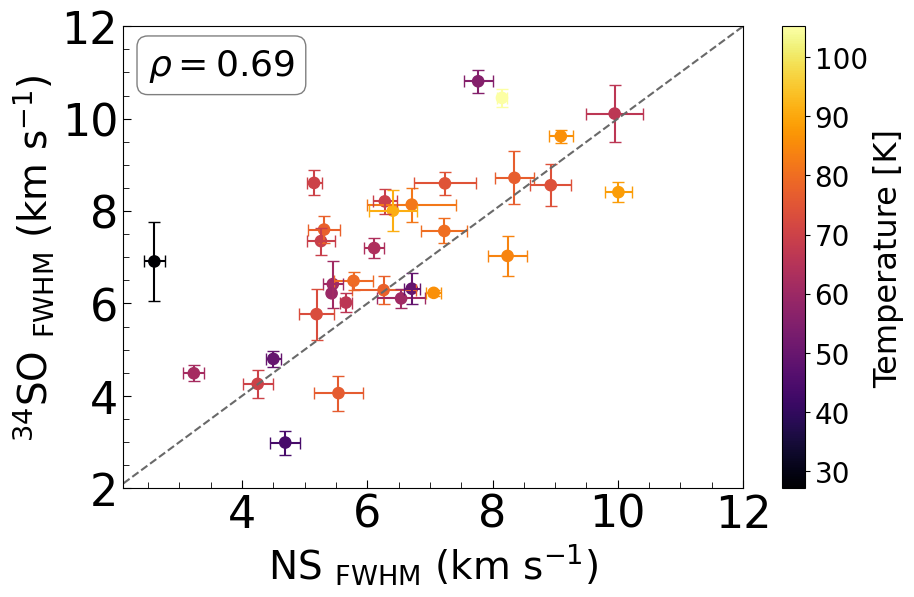}    
    \caption{Comparison of the $\Delta$v FWHM of the molecular lines. The points are color-coded based on the kinetic temperature measured in the cores. The dashed gray line represents the $y=x$ locus. The Pearson correlation coefficient ($\rho$) is displayed in each panel.}
    \label{DV}
\end{figure*}

In Fig.\,\ref{DV} we present the comparisons among all the line widths of the analyzed molecular species with a color code representing temperature. To better discern whether the emission of some molecular species is wider, or not, respect the other, it is displayed the unity line, and the Pearson coefficient is included in each panel to evaluate the correlation among the line width points. 

\section{Discussion}

Firstly, we discuss the results obtained from the comparisons among the molecular abundances, column densities, and physical parameters such as the kinetic temperature and the volume density.

From the relation between the abundances and temperature (see Fig.\,\ref{XvsT}), it can be noticed that the abundance of all analyzed sulfur-bearing species increases with the temperature, i.e. abundances are positively correlated with T$_{k}$. This is in agreement with what was found by \citet{fontani23} for the SO, SO$_{2}$, SO$^{+}$, and 
H$_{2}$CS in a sample of high-mass starless cores, high-mass protostellar objects, and ultracompact \hii~regions that were observed with single-dish telescope. 

Based on the linear fittings (see the obtained slopes in Fig.\,\ref{XvsT}), the increment of the molecular abundances with  T$_{k}$ is almost the same for all cases. Thus, we suggest that the chemistry involved in each of the analyzed sulfur-bearing species may have a similar dependence with T$_{k}$. This result, obtained for a temperature range going from about 20 to 100 K, supports the hypothesis that sulfur may be frozen in the dust grains in dense and cold interstellar regions \citep{laas19}, and hence, as the temperature rises, the abundance of sulfur-containing molecules also increases. Of course, we can not discard that the sulfur is already in the gaseous phase and the temperature increase activates chemical reactions favoring the production of such sulfur-bearing species.
It is worth noting that in the case of H$_{2}$CS, the obtained relation is very similar to what was obtained by \citet{chen24} from a large sample of massive protoclusters with diverse physical conditions, but in general, warmer than the cores investigated here.
Regarding the density of the cores, it seems that it has not any influence in the abundance-temperature relation, at least in the density range estimated here, coarsely between 10$^{6}$ and 10$^{7}$ cm$^{-3}$. 

Relations between the abundance of a molecular species and its column density (Fig.\,\ref{XvsNx}) show an interesting result. In all cases, the abundance increases with the column density, which indicates that the column density of the sulfur-containing molecule increases greater than the N(H$_{2}$). However, we can not rule out a possible underestimation in the  N(H$_{2}$) due to dust opacity effects. On the other hand, it is worth noting that in the case of thioformaldehyde, the relation obtained from the linear fitting is almost the same as that obtained by \citet{chen24} in their sample of massive protoclusters.  

Additionally, from the panels displayed in Fig.\,\ref{XvsNx}, it is clear that the temperature plays an important role in the increasing of the abundances; warm cores (orange and yellow points) are in general over the fitted lines in all cases. This indicates that warm cores have higher molecular abundances than cold ones at the same column density. Our results show that what was found by \citet{chen24} for the thioformaldehyde also applies to the other sulfur-bearing species, which could be an important point to be taken into account in chemical models studying the formation of such molecules.

\subsection{Comparisons among molecular abundances}

As presented in Fig.\,\ref{XvsX}, all comparisons among the abundances of the analyzed sulfur-bearing species present a positive correlation. Additionally, it is important to note that the increase in the X vs. X correlates with the kinetic temperature. This phenomenon shows that as temperature increases, abundances also increase, giving support, as mentioned above, to the hypothesis of the sulfur trapped in
dust-grain ice mantles in the cold regions. Moreover, based on the good, and in some cases very good, correlation among the abundances (coefficient of determination R$^{2}$ larger than 0.75, and in some cases about 0.9 in the linear regressions), we can conclude that the chemical processes involved in the formation of all analyzed sulfur-bearing species should be related. 

Interestingly, the correlation between X(SO$^{+}$) and X(SO) is the lowest among all pairs of compared abundances. Thus, taking into account that a possible route of formation of the SO$^{+}$ in this kind of region is through the SO ionization via cosmic rays \citep{rivilla22}, it is likely that as SO abundance increases, some of these molecules could be contributing to generating SO$^{+}$, which would be reflected in a loss of correlation between the abundances of both molecules. Moreover, the fact that the average $\Delta$v of both molecules are the same (see Fig.\,\ref{histo_DF}) suggests that both species coincide in the same gas layer.

\citet{fontani23} presented relations of abundances of SO$^{+}$, NS, SO$_{2}$, and H$_{2}$CS with SO, obtaining Pearson correlation coefficients ($\rho$) of: 0.62, 0.70, 0.95, and 0.73, respectively. In our case, the corresponding Pearson coefficients ($\rho = \sqrt{\rm R^{2}}$, being R$^{2}$ the determination coefficient of the line regressions presented in Fig.\,\ref{XvsX}) for such abundance relations are 0.83, 0.87, 0.93, and 0.93, respectively. All are larger than those presented by the mentioned authors, except $\rho$ of the  X(SO$_{2}$)/X(SO) relation, in this case, both values are almost the same. The fact that our sample contains sources at earlier evolutionary stages than the sample studied by \citet{fontani23} could explain the better correlation among the abundances found in our work. This could indicate that these molecules can have a similar chemical origin at these early stages (for instance the sulfur freeze-out from the dust grains), and then, such correlations among abundances would decrease as the cores evolve due to that the chemistry becomes more complex. 

Finally, it is worth noting that the obtained values of X(SO$_{2}$)/X(SO) ratio for our sample of cores are in agreement
with what is yielded by chemical models of a core chemically evolved about 10$^{5}$ yr for a temperature of 60 K (see \citealt{wak11}). The average measured temperature of our sample is 65 K. Even though we cannot know with accuracy the chemical ages of the cores in the analyzed sample, we can suggest that our observational results may support the use of this abundance ratio as a chemical clock as it was proposed by the authors.

\subsection{Line widths comparisons}

The obtained $\Delta$v values for the studied sulfur-bearing molecules are quite similar to those obtained, also from interferometric observations of sulfur species, towards others molecular cores \citep{elakel22,chen24}. It is worth noting that in this large sample of sources we must be careful with comparisons with other sources studied in the literature at different stages of evolution. The clumps in which the analyzed cores are embedded are cataloged as early star-forming regions. However, it is likely that this sample of clumps is not fully uniform, potentially exhibiting variations in evolutionary stage. As mention above, the $\Delta$v (FWHM) can be used to analyze the location or distribution of the molecular species in different gaseous layers with different dynamics within the cores \citep{sanhueza12,naiping21,fontani23,martinez24}.

Figure\,\ref{DV} shows that the line widths $\Delta$v (FWHM) of all molecular species are positively correlated, and although not so well defined, such a correlation is in general accompanied by the temperature increase. It is worth noting that none of the displayed $\Delta$v comparisons are close to the $y = x$ locus, only the relation between the line widths of $^{34}$SO and NS presents some approach to the unity line. This behavior indicates that, in general, the different molecules trace gas with different dynamics or turbulence conditions.

In particular, there is an excellent correlation between H$_2$CS and NS ($\rho = 0.93$), but the NS line widths are systematically larger. Assuming that the $\Delta$v in our sample of cores mainly reflects turbulence, the above-mentioned would indicate that NS traces the same gaseous layer as H$_2$CS but also has some contribution from more turbulent material. This is in agreement with the fact that the correlation between the abundances of these two molecules is among the highest in the sample ($\rho = 0.97$; see Fig.\,\ref{XvsX}).
The correlations of $\Delta$v of H$_2$CS and NS with the O-bearing species are generally poor ($\rho < 0.7$), except for the relation between NS and SO, which have $\rho = 0.78$. We note that the four O-bearing species have larger $\Delta$v than the other molecular species, indicating that they may be associated with warmer and more turbulent gas.
From Figs.\,\ref{histo_DF} and\,\ref{DV} it is clear that the oxygen-containing species trace the most turbulent gas. However, the correlations among them are quite poor, indicating that they may arise from different gaseous layers in the cores. In particular, SO$_2$ is probably located in the innermost region, which is more affected by a possible incipient central compact source, while the SO and $^{34}$SO could be located farther away. The nature of SO$^{+}$ is elusive. It presents the poorest correlation with the remaining oxygen-containing species. Besides, its $\Delta$v are generally larger than those of $^{34}$SO, but this is not evident when comparing with SO and SO$_2$. 

We suggest that H$_2$CS and NS are associated with quiescent gas, likely from the outer envelope of the cores, being NS probably located inner concerning H$_2$CS. The comparison between the FWHM $\Delta$v of these species presents the best correlation, which coincides with the high correlation in the comparison between their abundances. 
This indicates that the physics and chemistry that both molecular species may trace should be similar.

The oxygen-containing species may trace inner gas and are probably more affected by incipient star-forming processes. In particular, SO$_2$ traces the warmer and innermost regions.
It is worth noting that our results are in good agreement with those obtained by \citet{fontani23}, who found that the C- and N- bearing species are associated with the quiescent gas in the extended envelope of the sources while O-bearing species are located in inner regions of the cores.  The phenomenon that S and O bearing molecules are associated with more turbulent gas was also observed by other authors (see e.g. \citealt{artur19,garufi22,artur23}).

\section{Summary and conclusions}

We carried on an investigation of the sulfur-bearing molecules $^{34}$SO, SO$_{2}$, NS, SO, SO$^{+}$, and H$_{2}$CS towards  37 molecular cores embedded in a sample of infrared-quiet ATLASGAL sources presumably at very early stages of star formation. From the submillimeter continuum emission and methanol lines the cores were characterized in density and temperature, and the corresponding H$_{2}$ column densities were derived. The high rate of detection of the mentioned sulfur-bearing species allowed us to perform a thoughtful statistical study, useful for the investigation of the early sulfur chemistry in dense regions of the ISM. The main results and conclusions are summarized as follows:

1) It was observed that all the estimated abundances of the analyzed sulfur-bearing species increase, in a similar way, with the growth of the gas temperature. This supports the hypothesis that sulfur may be frozen in the dust grains in dense and cold interstellar regions, and hence, as the temperature rises, the abundance of sulfur-containing molecules also increases. It cannot be discarded that the sulfur is already present in the gaseous phase and the temperature increase activates some new chemical routes of formation of such sulfur-bearing species, increasing their abundances.
Additionally, we suggest that the chemistry involved in the formation of each of the analyzed sulfur-bearing species may have a similar dependence with T$_{k}$, at least in the range from 20 to 100 K. 

2) We find that the warmer cores of our sample have higher molecular abundances than the colder ones at the same column density for all molecular species. This complements a previous study done only for the thioformaldehyde, indicating that the mentioned behavior between abundances and column densities also applies to the other sulfur-bearing species. This could be a point to be taken into account in chemical models studying the formation of such molecules.

3) The molecular abundance comparisons present, in general, high correlations. If we compare with previous similar works done towards more evolved sources, we can conclude that the sulfur-bearing species here analyzed should have the same chemical origin at the very early stages of the cores, explaining the high correlations. Then, as the cores evolve, the correlations among the molecular abundances decrease, indicating that chemistry becomes more complex.

4) From the analysis of the FWHM $\Delta$v of the molecular lines, we suggest that the analyzed molecules with oxygen content ($^{34}$SO, SO, SO$^{+}$, and SO$_{2}$) may be associated with warmer and more turbulent gas than the other ones.
Our results indicate that the H$_{2}$CS and NS are associated with more quiescent gas, probably in the external envelopes of the cores, while the O-bearing species should be located in inner regions.   

5) In particular, NS and H$_2$CS show a very high correlation at both, the FWHM $\Delta$v and abundances comparisons. This may indicate that the physics and chemistry that both molecular species trace should be similar.

Finally, we point out that we are presenting a work with an important piece of information regarding the sulfur chemistry at the earliest stages of the molecular cores, probably pre-stellar cores. This work not only complement recent previous similar works done towards more evolved sources, in addition, it gives quantitative information about abundances that could be useful in chemical models pointing to explain the sulfur chemistry in the interstellar medium.

\begin{acknowledgements}

We would like to thank to the anonymous referee for her/his useful suggestions and corrections that improved our manuscript. N.C.M. is a doctoral fellow of CONICET. M.O., A.P., and S.P. are members of the {\it Carrera del Investigador Cient\'\i fico} of CONICET, Argentina.  A.A., M.B., C.C., and T.H. are part of our citizen science project {\it Ciencia Popular}\footnote{https://interestelariafe.wixsite.com/mediointerestelar/cienciapop}. This work was partially supported by the Argentina grants PIP 2021 11220200100012 and PICT 2021-GRF-TII-00061 awarded by CONICET and ANPCYT. 
This paper makes use of the following ALMA data: ADS/JAO.ALMA$\#$2017.1.00914.S ALMA is a partnership of ESO (representing its member states), NSF (USA) and NINS (Japan), together with NRC (Canada), MOST and ASIAA (Taiwan), and KASI (Republic of Korea), in cooperation with the Republic of Chile. The Joint ALMA Observatory is operated by ESO, AUI/NRAO and NAOJ. The National Radio Astronomy Observatory is a facility of the National Science Foundation operated under cooperative agreement by Associated Universities, Inc.

\end{acknowledgements}

%
%

\bibliographystyle{aa}  
\bibliography{ref}
\IfFileExists{\jobname.bbl}{}
{\typeout{}
\typeout{****************************************************}
\typeout{****************************************************}
\typeout{** Please run "bibtex \jobname" to optain}
\typeout{** the bibliography and then re-run LaTeX}
\typeout{** twice to fix the references!}
\typeout{****************************************************}
\typeout{****************************************************}
\typeout{}
}
\label{lastpage}

\begin{appendix} 

\section{Measured parameters from the molecular lines}
\label{append1}

Tables\,\ref{param1} and \ref{param2} present the line parameters, Peak, $\Delta$v and the integrated emission ($W$), obtained from each analyzed molecular lines from Gaussian fittings. For an example of such fittings see Appendix\,\ref{example}.

\begin{landscape}
\begin{table}[h]
\caption{Parameters obtained from Gaussian fittings.}
\label{param1}
\small
\begin{tabular}{l|cccccc|cccccc|cccccccccccccccccccccccc}
\hline
 & \multicolumn{6}{|c|}{SO$^{+}$ 15/2--13/2 (1/2) l=f } &   \multicolumn{6}{|c|}{H$_{2}$CS 10(1,9)--9(1,8) } & \multicolumn{6}{|c}{NS 15/2--13/2} \\
 \hline
Core &Peak&Error&$\Delta$v &Error&$W$&Error&Peak&Error&$\Delta$v&Error&$W$&Error&Peak&Error&$\Delta$v&Error&$W$&Error \\
&\multicolumn{2}{c}{\tiny(Jy beam$^{-1}$)}&\multicolumn{2}{c}{\tiny(km s$^{-1}$)} &\multicolumn{2}{c|}{\tiny(Jy beam$^{-1}$ km s$^{-1}$)} &\multicolumn{2}{c}{\tiny(Jy beam$^{-1}$)}&\multicolumn{2}{c}{\tiny(km s$^{-1}$)} &\multicolumn{2}{c|}{\tiny(Jy beam$^{-1}$ km s$^{-1}$)} & \multicolumn{2}{c}{\tiny(Jy beam$^{-1}$)}&\multicolumn{2}{c}{\tiny(km s$^{-1}$)} &\multicolumn{2}{c}{\tiny(Jy beam$^{-1}$ km s$^{-1}$)} \\
\hline
1&-&-&-&-&-&-&0.93&0.02&7.34&0.17&7.34&0.15&0.34&0.01&8.93&0.33&3.28&0.11 \\
2&0.12&0.01&6.32&0.31&0.86&0.04&3.58&0.10&4.46&0.15&17.07&0.54&1.97&0.04&6.11&0.16&12.85&0.32 \\
3&0.08&0.01&5.83&0.35&0.53&0.02&1.71&0.07&4.90&0.24&8.92&0.41&0.76&0.03&5.78&0.31&4.73&0.23 \\
4&0.09&0.01&6.47&0.48&0.61&0.03&1.68&0.02&6.76&0.09&12.15&0.15&1.36&0.02&9.09&0.19&13.25&0.25 \\
5&0.09&0.01&1.99&0.37&0.19&0.03&0.65&0.03&3.01&0.20&2.11&0.13&0.10&0.01&2.60&0.17&2.29&0.01 \\
6&0.17&0.01&9.63&0.47&1.77&0.07&2.95&0.1&6.17&0.25&19.41&0.73&2.06&0.06&8.15&0.28&17.88&0.57 \\
7&0.88&0.01&5.40&0.14&5.11&0.11&7.92&0.19&4.98&0.16&42.03&1.20&4.15&0.04&6.26&0.08&27.72&0.35 \\
8&0.06&0.01&8.99&0.82&0.59&0.04&1.51&0.06&4.52&0.20&7.28&0.31&0.22&0.02&5.15&0.51&1.24&0.11 \\
9&0.15&0.01&10.77&0.45&1.77&0.06&5.14&0.10&4.70&0.10&25.79&0.52&1.10&0.02&5.45&0.12&6.35&0.13 \\
10&0.04&<0.01&6.50&0.74&0.29&0.03&0.62&0.02&3.45&0.12&2.28&0.08&0.22&0.01&3.23&0.16&0.76&0.03 \\
11&0.19&0.01&6.79&0.16&1.37&0.03&5.02&0.04&5.38&0.06&28.81&0.30&1.81&0.04&5.66&0.17&10.93&0.31 \\
12&0.05&0.01&4.48&0.41&0.24&0.02&1.55&0.01&3.75&0.05&6.22&0.08&0.64&0.01&4.50&0.10&3.10&0.06 \\
13&0.13&0.01&7.70&0.20&1.13&0.02&2.30&0.03&4.48&0.09&11.90&0.21&0.29&0.01&6.54&0.12&2.04&0.03 \\
14&0.33&0.02&8.38&0.62&3.02&0.20&5.39&0.20&8.44&0.37&48.50&2.00&2.63&0.07&9.95&0.38&27.91&0.94 \\
15&-&-&-&-&-&-&1.84&0.08&5.42&0.27&10.67&0.50&0.43&0.02&6.71&0.45&3.07&0.18 \\
16&-&-&-&-&-&-&1.13&0.01&5.80&0.09&7.03&0.10&0.45&0.04&7.24&0.71&3.47&0.33 \\
17&0.29&0.01&8.07&0.31&2.55&0.09&2.31&0.06&5.51&0.17&13.55&0.41&0.38&0.01&7.06&0.49&2.91&0.17 \\
18&0.26&0.01&5.06&0.16&1.38&0.04&2.03&0.05&4.15&0.12&8.96&0.25&0.84&0.02&4.25&0.12&3.83&0.11 \\
19&0.11&0.01&5.85&0.37&0.68&0.04&2.46&0.05&6.07&0.16&15.92&0.39&0.77&0.02&6.71&0.24&5.50&0.18 \\
20&2.45&0.08&8.73&0.38&22.77&0.90&9.59&0.16&7.69&0.15&78.60&1.45&5.02&0.07&8.35&0.13&44.65&0.68 \\
21&-&-&-&-&-&-&1.66&0.04&4.85&0.13&8.61&0.22&0.60&0.02&6.28&0.31&4.07&0.19 \\
22&0.30&0.01&7.70&0.22&2.49&0.06&6.58&0.11&5.62&0.10&39.42&0.71&3.25&0.11&5.19&0.19&17.99&0.64 \\
23&0.61&0.03&8.58&0.57&5.60&0.33&7.36&0.17&6.47&0.17&50.75&1.28&1.47&0.04&7.77&0.28&12.22&0.39 \\
24&-&-&-&-&-&-&2.50&0.11&3.38&0.18&9.01&0.46&0.48&0.02&4.69&0.23&2.42&0.11 \\
25&-&-&-&-&-&-&0.37&0.03&3.14&0.36&1.25&0.13&-&-&-&-&-&- \\
26&-&-&-&-&-&-&0.89&0.02&5.06&0.15&4.82&0.14&0.53&0.02&5.54&0.24&3.17&0.13 \\
27&-&-&-&-&-&-&0.72&0.04&2.87&0.19&2.21&0.14&0.22&0.01&3.56&0.28&0.86&0.06 \\
28&0.06&0.01&8.47&1.21&0.59&0.07&1.89&0.07&4.61&0.22&9.30&0.41&0.55&0.03&5.31&0.39&3.10&0.21 \\
29&0.72&0.01&8.36&0.24&6.41&0.16&23.56&0.40&5.89&0.11&147.99&2.75&12.98&0.40&7.23&0.26&99.94&3.41 \\
30&2.86&0.12&8.37&0.40&25.52&1.14&26.14&0.89&5.44&0.21&151.53&5.58&16.08&0.62&8.24&0.37&141.24&5.92 \\
31&0.09&0.01&8.31&0.75&0.85&0.06&4.02&0.15&5.06&0.22&21.69&0.88&1.67&0.08&5.26&0.31&9.38&0.52 \\
32&0.91&0.03&9.42&0.38&9.09&0.34&9.94&0.19&5.66&0.12&60.02&1.27&7.08&0.21&6.41&0.22&48.38&1.61 \\
33&-&-&-&-&-&-&0.83&0.02&7.49&0.28&6.65&0.23&0.47&0.01&10.01&0.39&5.05&0.18 \\
34&-&-&-&-&-&-&0.04&0.01&4.13&0.96&0.19&0.04&-&-&-&-&-&- \\
35&-&-&-&-&-&-&-&-&-&-&-&-&-&-&-&-&-&- \\
36&-&-&-&-&-&-&0.20&0.02&3.15&0.38&0.66&0.07&0.11&0.01&2.10&0.25&0.25&0.02 \\
37&0.17&0.01&9.24&0.75&1.71&0.12&6.82&0.19&4.56&0.15&33.15&1.02&2.52&0.08&5.43&0.21&14.56&0.52 \\
\hline
\multicolumn{9}{l}{A `-' means that it was not observed emission from such molecular line.} \\ 
\end{tabular}
\end{table}
\end{landscape}

\begin{landscape}
\begin{table}[h]
\caption{Parameters obtained from Gaussian fittings.}
\label{param2}
\small
\begin{tabular}{l|cccccc|cccccc|cccccccccccccccccccccccc}
\hline
 & \multicolumn{6}{|c|}{SO v=0 $^{3}$$\Sigma$ 9(8)--8(7)} &   \multicolumn{6}{|c|}{$^{34}$SO 7(8)--6(7)} & \multicolumn{6}{|c}{SO$_{2}$ v=0 8(2,6)--7(1,7)} \\
 \hline
Core &Peak&Error&$\Delta$v &Error&$W$&Error&Peak&Error&$\Delta$v&Error&$W$&Error&Peak&Error&$\Delta$v&Error&$W$&Error \\
&\multicolumn{2}{c}{\tiny(Jy beam$^{-1}$)}&\multicolumn{2}{c}{\tiny(km s$^{-1}$)} &\multicolumn{2}{c|}{\tiny(Jy beam$^{-1}$ km s$^{-1}$)} &\multicolumn{2}{c}{\tiny(Jy beam$^{-1}$)}&\multicolumn{2}{c}{\tiny(km s$^{-1}$)} &\multicolumn{2}{c|}{\tiny(Jy beam$^{-1}$ km s$^{-1}$)} 
& \multicolumn{2}{c}{\tiny(Jy beam$^{-1}$)}&\multicolumn{2}{c}{\tiny(km s$^{-1}$)} &\multicolumn{2}{c}{\tiny(Jy beam$^{-1}$ km s$^{-1}$)} \\
\hline
1&1.27&0.03&8.04&0.26&10.89&0.33&0.14&0.01&8.56&0.46&1.33&0.06&0.23&0.01&9.11&0.24&2.23&0.05 \\
2&4.41&0.19&6.98&0.34&32.81&1.52&0.58&0.01&7.20&0.22&4.48&0.13&0.61&0.01&7.20&0.22&5.31&0.11 \\
3&3.24&0.12&5.12&0.23&17.69&0.76&0.32&0.01&6.49&0.20&2.23&0.06&0.50&0.01&8.13&0.3&4.40&0.15 \\
4&3.23&0.04&10.38&0.19&35.78&0.61&0.37&0.01&9.62&0.14&3.81&0.05&0.72&0.02&11.46&0.39&8.88&0.27 \\
5&2.57&0.12&2.41&0.13&6.60&0.34&0.06&0.01&6.91&0.85&0.44&0.05&0.13&0.01&2.9&0.43&0.42&0.06 \\
6&3.37&0.11&10.49&0.40&37.77&1.36&0.67&0.01&10.45&0.24&7.55&0.16&0.86&0.02&18.02&0.58&16.59&0.50 \\
7&10.52&0.23&8.03&0.20&8.95&2.13&2.83&0.07&6.29&0.19&18.99&0.54&3.78&0.15&7.71&0.36&31.11&1.36 \\
8&2.33&0.11&9.48&0.52&23.58&1.20&0.17&0.01&8.61&0.3&1.64&0.05&0.26&0.01&10.9&0.44&03.06&0.11 \\
9&6.80&0.29&8.29&0.42&60.13&2.84&0.66&0.02&6.42&0.27&4.53&0.17&0.91&0.05&8.67&0.6&8.47&0.54 \\
10&1.99&0.08&3.93&0.19&8.35&0.39&0.07&0.01&4.49&0.51&0.37&0.03&0.17&0.01&4.53&0.15&0.85&0.02 \\
11&6.09&0.16&6.88&0.21&44.62&1.32&0.91&0.02&6.02&0.17&5.85&0.15&0.21&0.01&15.99&1.30&3.72&0.27 \\
12&2.39&0.06&4.30&0.13&10.97&0.31&0.06&$<$0.01&4.80&0.21&0.33&0.10&0.06&$<$0.01&3.99&0.22&0.25&0.01 \\
13&5.16&0.14&6.35&0.20&34.90&1.05&0.45&0.01&6.11&0.18&2.98&0.08&0.62&0.01&7.83&0.25&5.18&0.15 \\
14&6.65&0.15&11.43&0.33&81.02&2.13&0.62&0.01&10.11&0.20&6.74&0.12&0.92&0.02&18.61&0.6&18.33&0.55 \\
15&2.56&0.09&6.06&0.32&16.55&0.77&0.12&0.01&8.14&0.61&1.10&0.08&0.20&0.01&4.91&0.16&1.09&0.03 \\
16&1.07&0.02&6.13&0.20&7.00&0.19&0.09&$<$0.01&8.6&0.37&0.91&0.03&0.11&0.01&7.64&0.32&0.93&0.03 \\
17&7.26&0.26&8.49&0.35&65.73&2.57&1.21&0.04&6.23&0.25&7.98&0.31&2.28&0.09&7.54&0.34&18.35&0.79 \\
18&0.85&0.02&4.25&0.11&3.84&0.11&0.81&0.01&4.26&0.07&3.65&0.06&1.45&0.01&4.58&0.05&7.10&0.08 \\
19&5.36&0.21&7..08&0.32&40.46&1.73&0.29&0.01&6.32&0.31&1.97&0.08&0.39&0.01&6.72&0.36&2.81&0.14 \\
20&19.96&0.21&12.26&0.15&247.53&2.98&6.30&0.20&8.72&0.33&58.47&2.09&7.56&0.23&10.55&0.37&85.00&2.81 \\
21&1.49&0.06&10.79&0.59&17.20&0.87&0.18&0.01&8.21&0.57&1.59&0.09&0.29&0.01&17.21&0.66&5.49&0.19 \\
22&12.2&0.23&7.31&0.16&94.98&2.01&1.26&0.05&5.77&0.27&7.74&0.34&2.27&0.08&6.08&0.27&17.74&0.61 \\
23&6.18&0.18&11.32&0.41&74.57&2.52&0.66&0.02&10.81&0.55&7.61&0.33&0.81&0.05&22.04&1.69&18.95&1.33 \\
24&2.21&0.08&5.41&0.24&12.79&0.54&0.11&0.01&2.99&0.25&0.35&0.02&0.10&0.01&4.94&0.93&0.55&0.09 \\
25&0.39&0.02&11.15&0.76&4.67&0.29&-&-&-&-&-&-&-&-&-&-&-&- \\
26&1.45&0.04&5.58&0.22&8.63&0.22&0.14&0.01&4.06&0.26&0.62&0.03&-&-&-&-&-&- \\
27&0.88&0.05&4.96&0.36&4.68&0.32&-&-&-&-&-&-&0.07&0.01&5.56&0.97&0.44&0.07 \\
28&2.76&0.15&5.52&0.35&16.26&0.96&0.22&0.01&7.60&0.38&1.77&0.08&0.30&0.01&9.80&0.47&3.17&0.14 \\
29&26.52&0.95&13.46&0.56&380.97&14.78&5.68&0.20&7.57&0.30&45.79&1.74&6.83&0.25&12.99&0.56&94.52&3.81 \\
30&38.35&1.04&10.96&0.34&447.66&13.22&11.44&0.38&7.03&0.27&85.74&3.09&18.26&0.50&8.09&0.25&157.33&4.67 \\
31&3.78&0.20&7.37&0.46&29.73&1.75&0.43&0.02&7.35&0.43&3.40&0.18&0.56&0.03&13.39&0.83&8.03&0.46 \\
32&11.71&0.40&9.43&0.45&117.64&5.25&2.06&0.06&8.01&0.29&17.63&0.60&2.63&0.08&18.68&0.76&52.29&1.93 \\
33&1.24&0.04&9.53&0.36&12.59&0.44&0.17&0.01&8.41&0.45&1.48&0.07&0.21&0.01&14.08&0.75&3.25&0.26 \\
34&-&-&-&-&-&-&-&-&-&-&-&-&-&-&-&-&-&- \\
35&0.19&0.03&1.62&0.38&0.32&0.07&-&-&-&-&-&-&-&-&-&-&-&- \\
36&0.50&0.05&3.34&0.41&1.78&0.20&-&-&-&-&-&-&-&-&-&-&-&- \\
37&8.57&0.35&7.49&0.36&68.45&3.07&0.81&0.02&6.22&0.22&5.34&0.17&1.47&0.04&7.25&0.04&11.32&0.36 \\
\hline
\multicolumn{9}{l}{A `-' means that it was not observed emission from such molecular line.} \\ 
\end{tabular}
\end{table}
\end{landscape}

\section{Calculation of optical depths}
\label{tauapp}

In order to evaluate the optically thin assumption used to calculate the column densities, we compute the optical depth ($\tau$) of 
the molecular lines in all cores from the following equation (e.g. \citealt{mangum15}): 

\begin{equation}
{\rm \tau = - ln\left(1 - \frac{T_{peak}}{\phi~(J(T_{ex}) - J(T_{bg}))}\right)},   
\label{RD} 
\end{equation}

\noindent where T$_{\rm peak}$ is the peak of the emission (values in the `Peak' columns in Tables\,\ref{param1} and \ref{param2} converted to K), $\phi$ the beam filling factor assumed to be 1, and J(T) = h$\nu$/k (exp(h$\nu$/kT)$-$1)$^{-1}$ with T$_{\rm ex} =$ 20 K, and T$_{\rm bg} = 2.73$ K. The calculated values of $\tau$ are presented in Table\,\ref{tau}.

\begin{table}[h]
\tiny
\caption{Line optical depths. }
\label{tau}
\begin{tabular}{lcccccc}
\hline
\hline
Core & $\tau$(SO$^{+})$&$\tau$(H$_{2}$CS)&$\tau$(NS)&$\tau$(SO)&$\tau(^{34}$SO)&$\tau$(SO$_{2}$) \\
\hline
1&-&0.030&0.011&0.041&0.005&0.007\\
2&0.004&0.120&0.065&0.151&0.020&0.019\\
3&0.003&0.055&0.024&0.109&0.011&0.016\\
4&0.003&0.054&0.044&0.108&0.013&0.023\\
5&0.003&0.021&0.003&0.085&0.002&0.004\\
6&0.005&0.097&0.068&0.113&0.023&0.027\\
7&0.028&0.287&0.142&0.407&0.100&0.124\\
8&0.002&0.049&0.007&0.077&0.006&0.008\\
9&0.005&0.176&0.036&0.243&0.022&0.029\\
10&0.001&0.020&0.007&0.065&0.002&0.005\\
11&0.006&0.172&0.059&0.215&0.031&0.007\\
12&0.002&0.050&0.021&0.079&0.002&0.002\\
13&0.004&0.075&0.009&0.179&0.015&0.019\\
14&0.010&0.186&0.087&0.237&0.021&0.029\\
15&0.002&0.060&0.014&0.085&0.004&0.006\\
16&-&0.036&0.014&0.035&0.003&0.003\\
17&0.009&0.075&0.012&0.262&0.042&0.073\\
18&0.008&0.066&0.027&0.027&0.028&0.046\\
19&0.003&0.081&0.025&0.187&0.010&0.012\\
20&0.080&0.359&0.174&1.005&0.238&0.266\\
21&-&0.054&0.019&0.048&0.006&0.009\\
22&0.010&0.232&0.109&0.490&0.043&0.073\\
23&0.019&0.264&0.048&0.218&0.022&0.025\\
24&-&0.082&0.015&0.073&0.004&0.003\\
25&-&0.012&-&0.012&-&-\\
26&-&0.028&0.017&0.047&0.005&-\\
27&-&0.023&0.007&0.028&-&0.002\\
28&0.002&0.061&0.018&0.092&0.007&0.009\\
29&0.023&1.353&0.532&1.846&0.212&0.237\\
30&0.095&1.730&0.716&-&0.486&0.831\\
31&0.003&0.135&0.055&0.128&0.015&0.017\\
32&0.029&0.375&0.255&0.465&0.072&0.085\\
33&-&0.026&0.015&0.040&0.006&0.007\\
34&-&0.001&-&-&-&-\\
35&-&-&-&0.006&-&-\\
36&-&0.006&0.004&0.016&-&-\\
37&0.005&0.242&0.084&0.318&0.028&0.046\\
\hline
\end{tabular}
\end{table}

Almost all values of $\tau$ are $<<1$, showing that indeed the lines are optically thin. Some values slightly larger than the unity (see cores 29 and 30 for the H$_{2}$CS and SO) could affect to the calculated column density in a factor about 2 (using the optical depth correction of $\tau$/(1$-$exp($-\tau$)), which does not change any result.

\section{Calculation of T$_{\rm rot}$ from the CH$_{3}$OH}
\label{appmetanol}

Table\,\ref{metanol} presents the integrated emission from the CH$_{3}$OH 7(1,7)--6(1, 6)$++$ (at 335.582 GHz, E$_{\rm u} =$ 78.97 K) and 12(1,11)--12(0,12)$-+$ (at 336.865 GHz, E$_{\rm u} =$ 197.07 K) lines used to estimate temperatures. 
Such integrated intensities, obtained from Gaussian fittings (see Fig.\,\ref{C4} for an example), were used to construct rotational diagrams following \citet{goldsmith99}. By assuming LTE conditions, optically thin lines, and a beam filling factor equal to the unity, we can estimate the rotational temperatures (${\rm T_{rot}}$). This analysis is based on a derivation of the Boltzmann equation,

\begin{equation}
{\rm ln\left(\frac{N_u}{g_u}\right)={\rm ln}\left(\frac{N_{tot}}{Q_{rot}}\right)-\frac{E_u}{kT_{rot}}},   \label{RotDia} 
\end{equation}

\noindent where ${\rm N_u}$ represents the molecular column density of the upper level of the transition, ${\rm g_u}$  the total degeneracy of the upper level, ${\rm N_{tot}}$ the total column density of the molecule, ${\rm Q_{rot}}$ the rotational partition function, and k the Boltzmann constant. 

Following \citet{miao95}, for interferometric observations, the left-hand side of Eq.\,\ref{RotDia} can be estimated from:

\begin{equation}
{\rm ln\left(\frac{N_u}{g_u}\right)={\rm ln}\left(\frac{2.04 \times 10^{20}}{\theta_a \theta_b}\frac{\it W}{g_{k}g_{l}{\nu_0}^3S_{ul}{\mu_0}^2}\right)},   
\label{RD} 
\end{equation}

\begin{table}[h]
\centering
\caption{CH$_{3}$OH integrated emission lines.}
\label{metanol}
\small
\begin{tabular}{lcccc}
\hline
\hline
Core&W$_{1}^{a,\dagger}$&Error&W$_{2}^{b,\dagger}$&Error \\
\hline
1&14.09&0.26&11.76&0.31\\
2&18.73&0.47&12.18&0.43\\
3&13.62&0.50&12.75&0.43\\
4&18.28&0.14&19.34&0.14\\
5&4.55&0.38&0.24&0.10\\
6&34.62&0.92&47.05&0.91\\
7&57.75&1.20&52.64&0.84\\
8&11.66&0.44&8.98&0.35\\
9&35.38&0.62&21.23&0.41\\
10&1.77&0.06&1.09&0.03\\
11&40.94&0.50&28.45&0.38\\
12&5.67&0.13&2.14&0.03\\
13&21.23&0.34&12.71&0.22\\
14&78.71&1.82&54.57&1.48\\
15&34.83&0.85&12.66&0.47\\
16&10.11&0.15&8.66&0.24\\
17&30.99&0.45&32.62&0.35\\
18&10.34&0.12&7.97&0.09\\
19&34.73&0.46&13.18&0.26\\
20&109.57&1.02&97.94&0.89\\
21&15.15&0.31&11.53&0.25\\
22&36.73&0.78&30.23&0.87\\
23&95.06&1.31&46.55&1.04\\
24&26.66&0.86&7.46&0.21\\
25&7.60&0.29&2.62&0.10\\
26&6.22&0.32&5.50&0.12\\
27&7.69&0.24&1.67&0.05\\
28&11.46&0.42&10.35&0.45\\
29&225.25&3.70&215.42&3.70\\
30&157.87&5.33&159.87&6.58\\
31&42.85&1.92&32.37&1.17\\
32&91.56&2.01&104.36&2.72\\
33&11.82&0.46&12.94&0.29\\
34&30.50&0.12&0.51&0.05\\
35&0.58&0.04&-&-\\
36&4.02&0.32&1.15&0.05\\
37&43.38&1.47&24.77&0.79\\
\hline
\multicolumn{5}{l}{\small $^{a}$ W$_{1}$: flux of the 7(1,7)--6(1, 6)$++$ line. } \\ 
\multicolumn{5}{l}{\small $^{b}$ W$_{2}$: flux of the 12(1,11)--12(0,12)$-+$ line.} \\
\multicolumn{5}{l}{$\dagger$ in Jy beam$^{-1}$ \ks.}

\end{tabular}
\end{table}

\noindent where $\theta_a$ and $\theta_b$ are the major and minor axes of the clean beam, respectively, $W$ is the integrated intensity, ${\rm g_k}$ and ${\rm g_l}$ are the K-level and reduced nuclear spin degeneracies, ${\rm \nu_0}$ is the rest frequency of the line, ${\rm S_{ul}}$ is the line strength of the transition, and $\mu_0$ is the permanent dipole moment of the molecule.  The free parameters, (${\rm N_{tot}/Q_{rot}}$) and ${\rm T_{rot}}$ were determined by a linear fitting to Eq.\,\ref{RotDia}. The obtained 
${\rm T_{rot}}$ values are those included in Table\,\ref{sample} as T$_{\rm CH_{3}OH}$.

The ${\rm T_{rot}}$ was estimated by assuming optically thin lines (e.g. \citealt{vander2000}). However, we consider the possibility of high optical depths in the used CH$_{3}$OH lines.  Thus, following \citet{goldsmith99} and \citet{purcell2006}, we compute the opacity correction factors for the ${\rm N_u}$ of both lines. We observe that in the cases that it makes sense to apply such a correction, it modifies the ${\rm N_u}$ value in a similar way for both transitions, and therefore it does not affect the ${\rm T_{rot}}$ estimation.

\section{Example of spectra and Gaussian fittings}
\label{example}

\begin{figure*}
    \centering
        \includegraphics[width=0.7\textwidth]{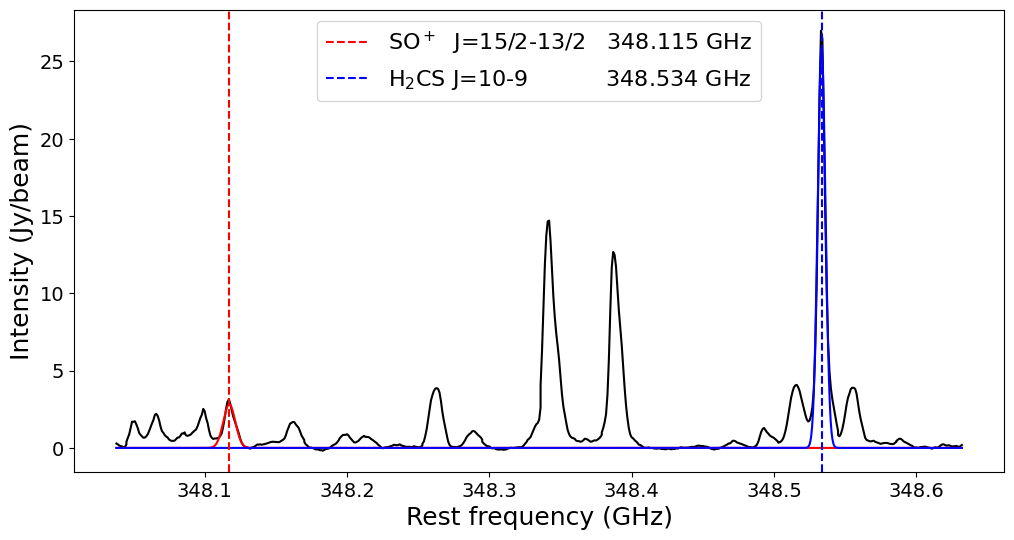}
        \caption{ALMA band 7 spw16 spectrum obtained towards the core in the ATLASGAL source 337.9154$-$0.4773. The analyzed SO$^+$ J=15/2$-$13/2 and H$_2$CS J=10$-$9 lines are indicated with their Gaussian fittings (red and blue, respectively).}
    \label{C.1.}
\end{figure*}

\begin{figure*}
    \centering
        \includegraphics[width=0.7\textwidth]{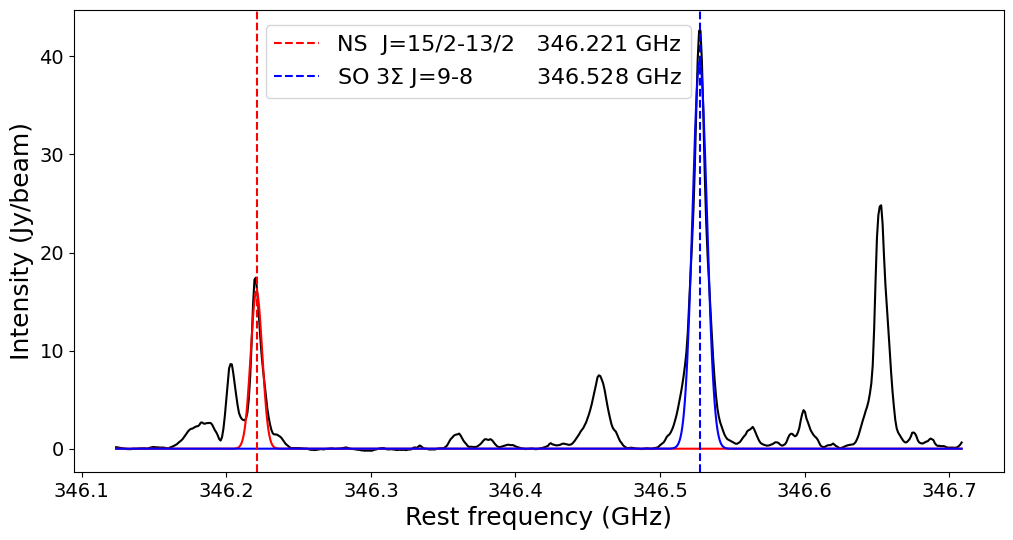}
        \caption{ALMA band 7 spw18 spectrum obtained towards the core in the ATLASGAL source 337.9154$-$0.4773. The analyzed NS J=15/2$-$13/2 and SO 3$\Sigma$ J=9$-$8 lines are indicated with their Gaussian fittings (red and blue, respectively).}
    \label{C.2.}
\end{figure*}

\begin{figure*}
    \centering
        \includegraphics[width=0.7\textwidth]{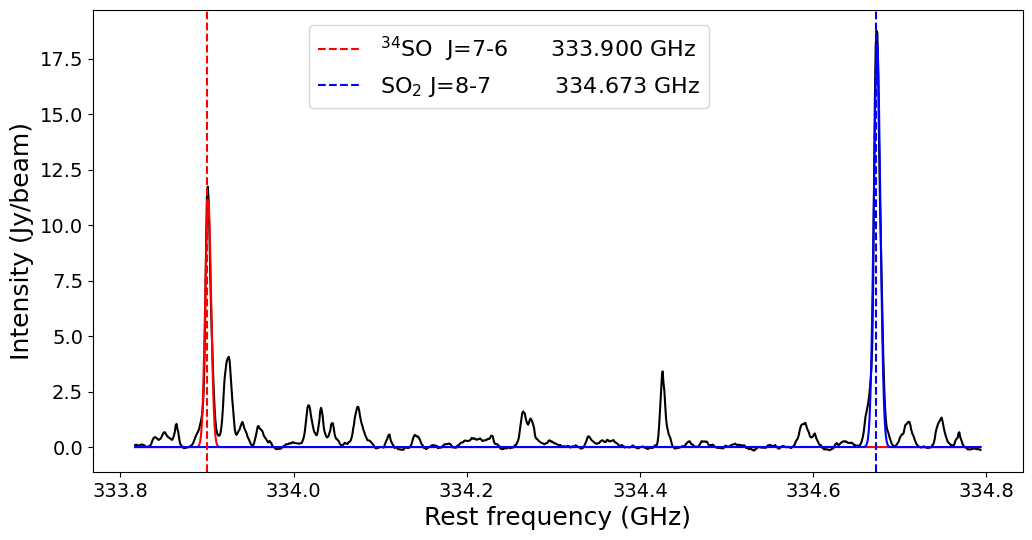}
        \caption{ALMA band 7 spw22 spectrum obtained towards the core in the ATLASGAL source 337.9154$-$0.4773. The analyzed $^{34}$SO J=7$-$6 and SO$_2$ J=8$-$7 lines are indicated with their Gaussian fittings (red and blue, respectively).}
    \label{C.3.}
\end{figure*}

\begin{figure*}
    \centering
        \includegraphics[width=0.7\textwidth]{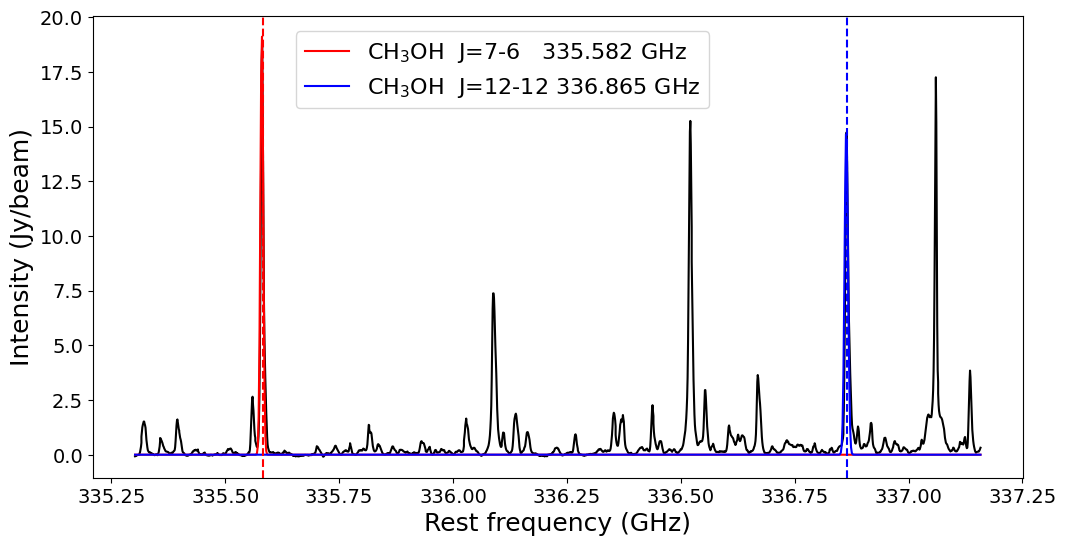}
        \caption{ALMA band 7 spw20 spectrum obtained towards the core in the ATLASGAL source 337.9154$-$0.4773. The used CH$_3$OH 7(1,7)--6(1, 6)$++$ and 12(1,11)--12(0,12)$-+$ lines to obtain the core temperature are indicated with their Gaussian fittings (red and blue, respectively).}
    \label{C4}
\end{figure*}

As an example, Figures \ref{C.1.}, \ref{C.2.}, and \ref{C.3.} display the spectra of band 7, spectral windows (spw) 16, 18, and 22, respectively, extracted from a beam centered at the position of the main core embedded in the ATLASGAL clump 337.9154$-$0.4773. The sulfur-bearing molecules are identified. The molecular line identification was made using CASA software by cross checking with the JPL and CDMS databases using the Splatalogue\footnote{https://splatalogue.online/\#/advanced}. Gaussian fittings are also included. Additionally, also as an example, Fig.\,\ref{C4} displays a band 7 spw 20 spectrum towards the same source showing the CH$_3$OH 7(1,7)--6(1, 6)$++$ and 12(1,11)--12(0,12)$-+$ lines used to obtain the core temperature. Gaussian fittings are also displayed.  

\end{appendix}

\end{document}